\documentclass{elsarticle}
\usepackage{lmodern}
\usepackage[T1]{fontenc}
\usepackage[latin9]{inputenc}
\usepackage{array}
\usepackage{verbatim}
\usepackage{url}
\usepackage{amsmath}
\usepackage{amssymb}
\usepackage{graphicx}
\PassOptionsToPackage{normalem}{ulem}
\usepackage{ulem}
\usepackage{longtable}
\usepackage{array}

\makeatletter
\usepackage{hyperref}
\usepackage{etoolbox}
\usepackage{color}

\usepackage{boxedminipage}

\usepackage{ifpdf}
\usepackage{ifthen}

\ifpdf
  \newcommand{\inputfig}[1]{
  }
\else
  \newcommand{\inputfig}[1]{%
\fi

\newcommand{\tsigma}{\mathcal{T}_{\Sigma}}
\newcommand{\tsigmav}{\tsigma(\Var)}
\newcommand{\crypt}[2]{\{#2\}_{#1}}
\newcommand{\sign}[2]{\mathsf{sign}_{#1}(#2)}
\newcommand{\scrypt}[2]{\{\!| #2 |\!\}_{#1}}
\newcommand{\inv}[1]{\iffont{inv}(#1)}
\newcommand{\invNA}{\iffont{inv}}
\newcommand{\pair}[2]{\langle #1, #2 \rangle}
\newcommand{\fst}[1]{\pi_1(#1)}
\newcommand{\snd}[1]{\pi_2(#1)}

\newcommand{\idfont}[1]{\mathit{#1}}
\newcommand{\HN}{\idfont{HN}}
\newcommand{\NA}{\idfont{NA}}
\newcommand{\NB}{\idfont{NB}}
\newcommand{\na}{\idfont{na}}
\newcommand{\nb}{\idfont{nb}}
\newcommand{\GX}{\idfont{GX}}
\newcommand{\GY}{\idfont{GY}}

\newcommand{\pkNA}{\iffont{pk}}
\newcommand{\skNA}{\iffont{sk}}
\newcommand{\pk}[1]{\pkNA(#1)}
\newcommand{\sk}[1]{\skNA(#1)}
\newcommand{\pkprime}[1]{\iffont{pk'}(#1)}

\newcommand{\senc}{\iffont{enc}}
\newcommand{\sdec}{\iffont{dec}}

\newcommand{\ckNA}{\mathsf{pk}}
\newcommand{\akNA}{\mathsf{sk}}
\newcommand{\ak}[1]{\akNA(#1)}
\newcommand{\ck}[1]{\ckNA(#1)}

\newcommand{\dyS}{\mathcal{DY}}
\newcommand{\dy}[1]{\dyS(#1)}
\newcommand{\dym}[2]{\dyS_{#1}(#2)}
\newcommand{\dymS}[1]{\dyS_{#1}}
\newcommand{\public}{\Sigma_p}

\newcommand{\state}[2]{\stateNA_{#1}(#2)}
\newcommand{\stateNA}{\iffont{state}}
\newcommand{\iknows}[1]{\iknowsNA(#1)}
\newcommand{\iknowsNA}{\iffont{ik}}
\newcommand{\ifnot}[1]{\iffont{not}(#1)}
\newcommand{\ifdot}{{}_{\;\bullet\;}}

\newcommand{\ifarrow}[1][]{
  \ifthenelse{\equal{#1}{}}{
    \Rightarrow
  }{
    =\!\!\![{#1}]\!\!\!\hspace{1pt}\Rightarrow
  }}

\newcommand{\iffont}[1]{\mathsf{#1}}

\newcommand{\roleA}{\iffont{roleA}} %
\newcommand{\roleB}{\iffont{roleB}} %
\newcommand{\roleR}{\iffont{roleR}} %
\newcommand{\roleC}{\iffont{roleC}} %

\newcommand{\secCh}{\mbox{$\,\bullet\!\!\rightarrow\!\!\bullet\,$}}
\newcommand{\athCh}{\mbox{$\,\bullet\!\!\rightarrow$\,}}
\newcommand{\cnfCh}{\mbox{${\rightarrow\!\!\bullet\,}$}}
\newcommand{\secRCh}{\,\bullet\!\!\!\twoheadrightarrow\!\!\bullet\,}
\newcommand{\athRCh}{\,\bullet\!\!\!\twoheadrightarrow}
\newcommand{\insecCh}{\rightarrow}

\newcommand{\fresh}[1]{\stackrel{\mbox{\scriptsize @}\,}{#1}}
\newcommand{\forward}{\looparrowright}
\newcommand{\fforward}{\fresh{\forward}}
\newcommand{\farrow}{\fresh{\rightarrow}}

\newcommand{\onionCh}{\,\bullet[\rightarrow]\bullet\,}

\newcommand{\idmxCh}{\onionCh}

\newcommand{\dotChdot}{\,\bullet\!\!\leftrightarrow\!\!\bullet\,}
\newcommand{\dotCh}{\,\bullet\!\!\leftrightarrow}
\newcommand{\Chdot}{\leftrightarrow\!\!\bullet\,}

\newcommand{\PsecChP}{\,\circ\!\!\!\rightarrow\!\!\circ\,}
\newcommand{\PsecCh}{\,\circ\!\!\!\rightarrow\!\!\bullet\,}
\newcommand{\secChP}{\,\bullet\!\!\!\rightarrow\!\!\circ\,}
\newcommand{\PathCh}{\,\circ\!\!\!\rightarrow}
\newcommand{\cnfChP}{\rightarrow\!\!\circ\,}

\newcommand{\PChP}{\,\circ\!\!\leftrightarrow\!\!\circ\,}
\newcommand{\dotChP}{\,\bullet\!\!\leftrightarrow\!\!\dot\,}
\newcommand{\PChdot}{\,\circ\!\!\leftrightarrow\!\!\bullet\,}
\newcommand{\PCh}{\,\circ\!\!\leftrightarrow}
\newcommand{\ChP}{\leftrightarrow\!\!\circ\,}

\newcommand{\FathChNA}{\iffont{athCh}}
\newcommand{\FcnfChNA}{\iffont{cnfCh}}
\newcommand{\FsecChNA}{\iffont{secCh}}
\newcommand{\FfathChNA}{\iffont{fathCh}}
\newcommand{\FfsecChNA}{\iffont{fsecCh}}
\newcommand{\FsecCh}[4]{\FsecChNA(#1;#2;#3;#4)}
\newcommand{\FathCh}[3]{\FathChNA(#1;#2;#3)}
\newcommand{\FcnfCh}[2]{\FcnfChNA(#1;#2)}
\newcommand{\FfathCh}[4]{\FfathChNA^{#1}(#2;#3;#4)}
\newcommand{\FfsecCh}[5]{\FfsecChNA^{#1}(#2;#3;#4;#5)}

\newcommand{\athIssue}[1]{\iffont{athIssue(#1)}}
\newcommand{\cnfIssue}[1]{\iffont{cnfIssue(#1)}}
\newcommand{\secIssue}[1]{\iffont{secIssue(#1)}}
\newcommand{\athRely}[1]{\iffont{athRely(#1)}}
\newcommand{\cnfRely}[1]{\iffont{cnfRely(#1)}}
\newcommand{\secRely}[1]{\iffont{secRely(#1)}}

\newcommand{\pkEnc}[1]{\mathit{pkEnc}(#1)}
\newcommand{\pkSig}[1]{\mathit{pkSig}(#1)}

\newcommand{\DownGrade}{\mathrm{DownGrade}}
\newcommand{\Combine}{\mathrm{Combine}}
\newcommand{\SymOne}{\mathrm{Sym_1}}
\newcommand{\SymTwo}{\mathrm{Sym_2}}
\newcommand{\SymTre}{\mathrm{Sym_3}}
\newcommand{\CreatePseudo}{\mathrm{CreatePseudo}}
\newcommand{\UsePseudo}{\mathrm{UsePseudo}}
\newcommand{\UseRealName}{\mathrm{UseRealName}}
\newcommand{\PseudoDownGrade}{\mathrm{PseudoDownGrade}}
\newcommand{\AuthPseudo}{\mathrm{AuthPseudo}}
\newcommand{\AthTTP}{\mathrm{AthTTP}}
\newcommand{\CnfTTP}{\mathrm{CnfTTP}}
\newcommand{\SecTTP}{\mathrm{SecTTP}}

\newcommand{\honest}[1]{\mathit{honest}(#1)}
\newcommand{\dishonest}[1]{\iffont{dishonest}(#1)}
\newcommand{\vddash}{\vdash\!\!\!\vdash}
\newcommand{\mmodels}{\models\hspace{-0.3cm}\models}

\newcommand{\mode}[3]{({#1} \,|\, {#2} \,|\, {#3})}

\newcommand{\IK}{\mathit{IK}}
\newcommand{\agent}[1]{\iffont{agent}(#1)}

\newcommand{\Fresh}{\mathit{Fresh}}
\newcommand{\Payload}{\mathit{Payload}}
\newcommand{\Public}{\mathit{Public}}
\newcommand{\Tag}{\mathit{Tag}}
\newcommand{\lift}[1]{\lceil #1\rceil}

\newcommand{\pos}[1]{\mathit{pos}(#1)}
\newcommand{\Var}{\mathcal{V}}

\newcommand{\tagfont}[1]{\mathsf{#1}}
\newcommand{\tSone}{\tagfont{S_1}}
\newcommand{\tStwo}{\tagfont{S_2}}

\newcommand{\optfix}[2]{}
\newcommand{\fix}[2]{{\bf FIX}\footnote{{\bf FIX #1}: #2}}
\newcommand{\dyM}[2]{\mathcal{DY}_{#1}(#2)}
\newcommand{\nf}[1]{#1_{\downarrow C/F}}
\newcommand{\pattern}[2]{{\lceil\,#1\,\rceil}_{#2}}
\newcommand{\decryptPat}[2]{{\lceil\!\!\lceil\,#1\,\rceil\!\!\rceil}_{#2}}

\newcommand{\pubNA}{\mathit{pub}}
\newcommand{\responseNA}{\mathit{response}}
\newcommand{\checkRNA}{\mathit{check}}

\newcommand{\pub}[1]{\pubNA(#1)}
\newcommand{\response}[1]{\responseNA(#1)}
\newcommand{\checkR}[1]{\checkRNA(#1)}

\newcommand{\Keyword}[1]{\mathsf{#1}}

\newcommand{\Protocol}[5]{
  \Keyword{Protocol:}~\mathit{#1}\\
  \Keyword{Types:}\\
  #2
  \Keyword{Knowledge:}\\
  #3
  \Keyword{Actions:}\\
  \begin{array}{cllllcl}
  #4
  \end{array}\\
  \Keyword{Goals:}\\
  \begin{array}{ccccl}
  #5
  \end{array}
}
\newcommand{\MidProtocol}[5]{
  \begin{array}{ccccl}
  #4
  \end{array}\\
}

\newcommand{\MSC}[5]{
  \Keyword{Protocol:}~\mathit{#1}\\
  \begin{array}{ccccl}
    #4
  \end{array}
}

\newcommand{\ShortProtocol}[5]{
  \Keyword{Protocol:}~\mathit{#1}\\
  \begin{array}{ccccl}
  #4
\end{array}\\
}
\newcommand{\CompactProtocol}[5]{
  \begin{array}{ccccl}
  #4
  \hline
  #5
  \end{array}\\
}
\newcommand{\Type}[2]{\quad #1\;\mathit{#2};\\}
\newcommand{\Agent}{\Keyword{Agent}}
\newcommand{\Number}{\Keyword{Number}}
\newcommand{\Function}{\Keyword{Function}}
\newcommand{\TFunction}{\Keyword{Function}}
\newcommand{\Knowledge}[2]{\quad\mathit{#1}:~\mathit{#2};\\}
\newcommand{\Create}[2]{
  \multicolumn{5}{l}{\quad\#\mathit{#1}~\text{creates}~\mathit{#2}}\\}
\newcommand{\Let}[2]{
  \multicolumn{5}{l}{\quad\#\mathit{#1}~:=~\mathit{#2}}\\}
\newcommand{\Action}[5]{\quad\mathit{#1}#2\mathit{#3},\mathit{#4}:&\mathit{#5}\\}
\newcommand{\Repeat}[5]{\multicolumn{5}{l}{
    \quad#1=\mathit{#2}#3\mathit{#4}:\mathit{#5}}\\}
\newcommand{\NGoal }[4]{\quad\mathit{#1}&#2&\mathit{#3}&:&\mathit{#4}\\}
\newcommand{\AuthGoal}[3]
{\multicolumn{5}{l}{
    \quad\mathit{#1}~\Keyword{authenticates}~\mathit{#2}~%
    \Keyword{on}~\mathit{#3}}\\}
\newcommand{\SecGoal}[2]
{\multicolumn{5}{l}{\quad\mathit{#1}~\Keyword{secret~between}~\mathit{#2}}\\}
\newcommand{\REML}[1]{\multicolumn{5}{l}{\quad\#\text{ #1}}\\}

\newcommand{\ifrule}[3]{#1\\ \ifarrow[#2]\\ #3\\[2ex]}

\newcommand{\subterm}{\sqsubseteq}
\newcommand{\propersubterm}{\sqsubset}
\newcommand{\supterm}{\sqsupseteq}
\newcommand{\propersupterm}{\sqsupset}

\newcommand{\decryptions}[2]{\mathit{decryptions}_{#1}(#2)}
\newcommand{\patternR}[3]{\pattern{#2}{#3}^{#1}}

\newcommand{\instanceof}{\succeq}

\newcommand{\secret}[2]{\iffont{secret}(#1,#2)}

\newcommand{\idemix}{\textsf{Identity Mixer}}

\newcommand{\xor}{\oplus}
\newcommand{\algo}[1]{\ensuremath{\mathsf{#1}}}
\newcommand{\const}[1]{\algo{#1}}
\newcommand{\vari}[1]{\ensuremath{\mathit{#1}}}

\newcommand{\ana}[2]{\mathit{ana}_{#1}(#2)}
\newcommand{\derivations}[3]{\mathit{derivations}_{#1}(#2,#3)}
\newcommand{\dereq}[3]{\mathit{dereq}_{#1}(#2,#3)}
\newcommand{\checks}[3]{\mathit{checks}_{#1}(#2,#3)}
\newcommand{\given}[1]{\textsl{given}(#1)}
\newcommand{\ungive}[1]{{#1}_*}

\definecolor{grey}{rgb}{0.5,0.5,0.5}\newcommand{\grey}{\color{grey}}
\definecolor{red}{rgb}{1,0,0}\newcommand{\red}{\color{red}}
\definecolor{green}{rgb}{0,0.45,0}\newcommand{\green}{\color{green}}
\definecolor{blue}{rgb}{0,0,1}\newcommand{\blue}{\color{blue}}

\newcommand{\rpif}{\mathbb{P}}
\newcommand{\rif}{\mathbb{R}}
\newcommand{\rf}{\mathbb{F}}
\newcommand{\re}{\mathbb{E}}
\newcommand{\md}{\mathbb{M}}
\newcommand{\traces}{\mathbb{T}}
\newcommand{\sigS}{\iffont{pk}(S)}
\newcommand{\send}[2]{\iffont{snd}_{#1}(#2)}
\newcommand{\recv}[2]{\iffont{rcv}_{#1}(#2)}
\newcommand{\trace}{t}
\newcommand{\evs}{\mathit{evs}}
\newcommand{\evt}{\mathit{evt}}
\newcommand{\mkset}[1]{[#1]} %
\newcommand{\player}[1]{\mathit{player}(#1)}
\newcommand{\used}[1]{{\mathit{ used\;#1}}}

\newcommand{\sem}[1]{[\![ #1 ]\!]}

\newcommand{\mX}{\mathcal{X}}

\newcommand{\PVar}{\mathcal{P}}
\newcommand{\SigmaP}{\Sigma_\PVar}
\newcommand{\tsigmap}{\mathcal{T}_{\SigmaP}}
\newcommand{\arity}[1]{\mathit{arity}(#1)}

\newcommand{\Nat}{\mathbb{N}}
\newcommand{\dcrypt}[2]{\crypt{#1}{#2}^{-1}}
\newcommand{\dscrypt}[2]{\scrypt{#1}{#2}^{-1}}
\newcommand{\ccs}[1]{\mathit{ccs}(#1)}

\newcommand{\interpretation}{\mathcal{I}}
\newcommand{\intruder}{\mathsf{i}}

\newcommand{\know}[1]{\mathit{know}(#1)}
\newcommand{\verify}[1]{\mathit{verify}(#1)}
\newcommand{\true}{\mathit{true}}

\newcommand{\anl}[1]{&&&&\%\mathit{#1}\\}
\newcommand{\REM}[1]{\;\;\Keyword{\#}\;\;\text{#1}}

\newcommand{\StoreOA}[2]{\Keyword{Store}_{#1}(\,#2\,)}
\newcommand{\CheckStoreOA}[2]{\Keyword{CheckStore}_{#1}(\,#2\,)}
\newcommand{\LoadOA}[2]{#2\leftarrow\Keyword{Load}_{#1}}

\newcommand{\Store}[2]{\multicolumn{5}{l}{
    \quad\StoreOA{#1}{#2}}\\}
\newcommand{\Load}[2]{\multicolumn{5}{l}{
    \quad\LoadOA{#1}{#2}}\\}
\newcommand{\CheckStore}[2]{\multicolumn{5}{l}{
    \quad\CheckStoreOA{#1}{#2}}\\}

\newcommand{\FormNym}{\mathsf{FormNym}}
\newcommand{\GrantCred}{\mathsf{GrantCred}}
\newcommand{\VerifyCred}{\mathsf{VerifyCred}}
\newcommand{\VerifyCredOnNym}{\mathsf{VerifyCredOnNym}}
\newcommand{\masecNA}{\iffont{masec}}
\newcommand{\masec}[1]{\masecNA(#1)}
\newcommand{\ptagNA}{\iffont{p}}
\newcommand{\ptag}[1]{\ptagNA(#1)}

\newcommand{\commitNA}{\iffont{commit}}
\newcommand{\commitIINA}{\iffont{commit_2}}
\newcommand{\commitIIINA}{\iffont{commit_3}}

\newcommand{\commit}[1]{\commitNA(#1)}
\newcommand{\commitII}[1]{\commitIINA(#1)}
\newcommand{\commitIII}[1]{\commitIIINA(#1)}

\newcommand{\zkpNA}{\iffont{zkp}}
\newcommand{\user}[1]{\iffont{user}(#1)}
\newcommand{\owner}[1]{\iffont{owner}(#1)}

\newcommand{\Always}{\iffont{Always}}
\newcommand{\Previously}{\iffont{Previously}}

\newcommand{\pending}[1]{\iffont{pending}(#1)}

\newcommand{\pack}[1]{\mathit{pack}(#1)}
\newcommand{\hide}[1]{\mathit{hide}(#1)}

\newcommand{\ToTrusted}[1]{\Action{#1}{\idmxCh}{T}}
\newcommand{\TrustedTo}[1]{\Action{T}{\idmxCh}{#1}}

\newcommand{\registered}[1]{\iffont{registered}(#1)}
\newcommand{\granted}[1]{\iffont{granted}(#1)}

\newcommand{\verified}[1]{\iffont{verified}(#1)}

\newcommand{\dhh}{\mathit{dhh}}
\newcommand{\dhf}{\mathit{dhf}}
\newcommand{\DHtag}[1]{\mathit{tag}(#1)}
\newcommand{\hk}{\mathit{hk}}
\newcommand{\fk}{\mathit{fk}}
\newcommand{\exps}{\mathit{exps}}

\newcommand{\from}[2]{\fromnoarg(#1;#2)}
\newcommand{\fromnoarg}{\mathit{from}}

\renewcommand{\vector}[1]{\overline{#1}}

\newcommand{\vars}[1]{\mathit{vars}(#1)}
\newcommand{\dom}[1]{\mathit{dom}(#1)}

\newcommand{\drightarrow}{\rightarrow\!\!\!\!\!\rightarrow}
\newcommand{\bigsem}[1]{\left[\!\!\left[#1\right]\!\!\right]}

\newcommand{\hornarrow}{\rightarrow}
\newcommand{\iknowsa}[1]{\mathsf{iknows}(#1}
\newcommand{\signa}[1]{\mathsf{sign}_{#1}}
\newcommand{\todo}[1]{\fbox{\textbf{TODO: }#1}}
\newcommand{\valid}{\mathit{valid}}
\newcommand{\revoked}{\textit{revoked}}
\newcommand{\db}{\mathit{db}}
\newcommand{\ring}{\mathit{ring}}
\newcommand{\PK}{\mathit{PK}}
\newcommand{\NPK}{\mathit{NPK}}
\newcommand{\occurs}{\mathit{occurs}}
\newcommand{\timplies}{\mathit{implies}}
\newcommand{\LFP}{\mathit{LFP}}

\newcommand{\Habs}{\hat{\mathfrak{A}}}
\newcommand{\Abs}{\mathfrak{A}}
\newcommand{\Vara}{\Var_{\mathfrak{A}}}
\newcommand{\tabs}{\mathcal{T}_\mathfrak{A}}
\newcommand{\thabs}{\hat{\mathcal{T}_\mathfrak{A}}}
\newcommand{\ITIF}{{\it IF}}
\newcommand{\CryptoIF}{{\it CCM}}
\newcommand{\ChanIF}{{\it ICM}}

\newcommand{\ann}{\textit{ann}}
\newcommand{\atag}{\mathsf{atag}}
\newcommand{\fatag}{\mathsf{fatag}}
\newcommand{\ctag}{\mathsf{ctag}}
\newcommand{\stag}{\mathsf{stag}}
\newcommand{\fstag}{\mathsf{fstag}}
\newcommand{\btag}{\mathsf{plain}}
\newcommand{\ftag}{\mathsf{fresh}}
\newcommand{\fotag}{\mathsf{forw}}
\newcommand{\blind}{\mathsf{blind}}
\newcommand{\verif}[1]{#1}
\newcommand{\seenNA}{\mathsf{seen}}
\newcommand{\seen}[1]{\seenNA(#1)}
\newcommand{\notseen}[1]{\mathsf{fresh}(#1)}

\newcommand{\ttbracket}[1]{\texttt{[}#1\texttt{]}}
\newcommand{\anb}{\textit{AnB}}
\newcommand{\anbx}{\textit{AnBx}}
\newcommand{\IM}{\texttt{IM}}
\newcommand{\CM}{\texttt{CM}}
\newcommand{\trans}[3]{\llbracket#1\rrbracket^{#2}_{#3}}
\newcommand{\enc}[1]{\llbracket #1 \rrbracket}
\newcommand{\itrans}[3]{\llparenthesis #1 \rrparenthesis^{#2}_{#3}}
\newcommand{\after}[2]{#1 \texttt{ after } #2}
\newcommand{\arr}[1]{=\!\![#1]\!\!\Rightarrow}
\newcommand{\x}{{\cal X}}
\newcommand{\y}{{\cal Y}}
\newcommand{\X}{{\cal X}}
\newcommand{\T}{{\cal T}}
\newcommand{\V}{{\cal V}}
\newcommand{\chan}{\textsf{chan}}

\newcommand{\cchannel}{\mathsf{Chan_{CCM}}}
\newcommand{\ichannel}{\mathsf{Chan_{ICM}}}
\newcommand{\mgu}{\textit{mgu}}
\newcommand{\bl}{\mathsf{b}}

\renewcommand{\sectionautorefname}{\S}
\renewcommand{\subsectionautorefname}{\S}
\newcommand{\Sectionautorefname}{\S}
\newcommand{\SubSectionautorefname}{\S}
\newcommand{\miexampleautorefname}{example}
\newcommand{\exampleautorefname}{example}
\renewcommand{\exp}{\iffont{exp}}
\renewcommand{\figureautorefname}{figure}

\renewcommand{\dscrypt}[2]{\scrypt{#1}{#2}}
\renewcommand{\dcrypt}[2]{\crypt{#1}{#2}}

\newcommand{\signed}[2]{\mathsf{S}_{#1}(#2)}
\newcommand{\nonce}[1]{n_{#1}}
\newcommand{\newkey}[1]{\mathsf{new}\enskip{#1}}
\newcommand{\hmac}[2]{\mathsf{hmac}_{#1}(#2)}
\newcommand{\hashsymbol}{\mathcal{H}}
\newcommand{\hash}[1]{\mathsf{hash}(#1)}
\newcommand{\verdigest}[2]{[\textit{#1:#2}]}
\newcommand{\digest}[1]{[\textit{#1}]}

\newcommand{\contains}[2]{\iffont{contains}(#1,#2)}
\newcommand{\error}{\iffont{error}}
\newcommand{\wrequest}[1]{\iffont{wrequest}(#1)}
\newcommand{\request}[1]{\iffont{request}(#1)}
\newcommand{\witness}[1]{\iffont{witness}(#1)}

\newcommand{\maxd}[1]{\mathit{maxd}(#1)}
\newcommand{\sccs}[1]{\mathit{sccs}(#1)}
\newcommand{\nts}[1]{\mathit{#1}}

\newcommand{\channel}[2]{\mathsf{channel}_{#1}(#2)}
\newcommand{\GYS}{\mathit{GYS}}

\newcommand{\ccsF}[1]{\mathit{ccs}_F(#1)}

\newcommand{\olds}[1]{\oldstylenums{#1}}
\newcommand{\oldsb}[1]{{\bfseries\olds{#1}}}
\newcommand{\mnote}[1]{\stepcounter{ncomm}%
\vbox to0pt{\vss\llap{\tiny\oldsb{\arabic{ncomm}}}\vskip6pt}%
\marginpar{\tiny\bf\raggedright%
{\oldsb{\arabic{ncomm}}}.\hskip0.5em#1}}\newcounter{ncomm}

\newcommand{\miknote}[1]{{\red \mnote{{\red m: #1}}}}

\usepackage{makeidx}%
\usepackage{array}
\usepackage{float}
\usepackage{multirow}
\usepackage{textcomp}
\usepackage{tikz}\usetikzlibrary{matrix,shapes,arrows,positioning,chains, calc}
\usepackage{algpseudocode}
\usepackage{algorithm}
\usepackage{url}
\usepackage{chngpage}

\providecommand{\tabularnewline}{\\}
\floatstyle{ruled}
\newfloat{algorithm}{tbp}{loa}
\providecommand{\algorithmname}{Algorithm}
\floatname{algorithm}{\protect\algorithmname}

\newcommand{\trNA}{\iffont{tr}}
\newcommand{\tr}[1]{\trNA(#1)}
\newcommand{\payreq}{\ensuremath\mathrm{\emph{PaymentRequest}}}
\newcommand{\paynow}{\ensuremath\mathrm{paynow}}
\newcommand{\refundAuth}{\ensuremath\mathrm{RA}}
\newcommand{\inputTransaction}{\ensuremath\mathrm{\pi}}
\newcommand{\bitcoinAddress}{\ensuremath\mathrm{B}}
\newcommand{\refundAddress}{\ensuremath\mathrm{R}}
\newcommand{\transactionid}{\ensuremath\mathrm{\tau'}}
\newcommand{\transaction}{\ensuremath\mathrm{\tau}}
\newcommand{\alice}{\ensuremath\mathrm{C}}
\newcommand{\bob}{\ensuremath\mathrm{M}}
\newcommand{\trader}{\ensuremath\mathrm{T}}
\newcommand{\caroline}{\ensuremath\mathrm{PP}}
\newcommand{\david}{\ensuremath\mathrm{D}}
\newcommand{\mallory}{\ensuremath\mathrm{i}}
\newcommand{\eve}{\ensuremath\mathrm{E}}
\newcommand{\zkp}{\ensuremath\mathrm{ZKP}}
\newcommand{\kdf}{\ensuremath\mathrm{KDF}}
\def\bitcoinB{\leavevmode
  {\setbox0=\hbox{\textsf{B}}%
    \dimen0\ht0 \advance\dimen0 0.2ex
    \ooalign{\hfil \box0\hfil\cr
      \hfil\vrule height \dimen0 depth.2ex\hfil\cr
    }%
  }%
}
\raggedbottom
\def\bitcoinC{\leavevmode\rlap{\hskip.5pt-}B}

\makeatother

\usepackage{listings}
\usepackage[scaled=0.85]{beramono}

\lstdefinelanguage{AnBx}{
    keywords = {Protocol:, Types:, Agent, Number, Function, Certified, SymmetricKey, PublicKey, Untyped, Definitions:, Knowledge:, where, agree, share, Actions:, Goals:, weakly, authenticates, on, secret, between, pk, empty},
    alsoletter=:,
    breaklines=true,
    basicstyle=\footnotesize\ttfamily,%
    keywordstyle=\bfseries,
    commentstyle=\color{green!40!black}\itshape,
    morecomment=[l]{\#}
}

\begin{document}

\title{Formal Modelling and Security Analysis\\ of Bitcoin's Payment Protocol\footnote{This is an accepted manuscript to appear in Computers \& Security. Please cite as: Modesti, Shahandashti, McCorry, and Hao. "Formal Modelling and Security Analysis of Bitcoin's Payment Protocol". To appear in Computer \& Security, Elsevier, 2021.}}

\author[1]{Paolo Modesti\corref{cor1}}
  \ead{p.modesti@tees.ac.uk}
\author[2]{Siamak F. Shahandashti}
  \ead{siamak.shahandashti@york.ac.uk}
\author[3]{Patrick McCorry}
\author[4]{Feng Hao}
  \ead{feng.hao@warwick.ac.uk}

\address[1]{Department of Computing and Games, Teesside University, UK}
\address[2]{Department of Computer Science, University of York, UK}
\address[3]{PISA Research, UK}
\address[4]{Department of Computer Science, University of Warwick, UK}

\cortext[cor1]{Corresponding author}

\begin{abstract}
The Payment Protocol standard BIP70, specifying how payments in Bitcoin are performed by merchants and customers, is supported by the largest payment processors and most widely-used wallets.
The protocol has been shown to be vulnerable to refund attacks due to lack of authentication of the refund addresses.
In this paper, we give the first formal model of the protocol and formalise the refund address security goals for the protocol, namely refund address authentication and secrecy.
The formal model utilises communication channels as abstractions conveying security goals on which the protocol modeller and verifier can rely.
We analyse the Payment Protocol confirming that it is vulnerable to an attack violating the refund address authentication security goal.
Moreover, we present a concrete protocol revision proposal supporting the merchant with publicly verifiable evidence that can mitigate the attack.
We verify that the revised protocol meets the security goals defined for the refund address.
Hence, we demonstrate that the revised protocol is secure, not only against the existing attacks, but also against any further attacks violating the formalised security goals.
\end{abstract}


\maketitle

\section{Introduction\label{sec:Introduction}}
Bitcoin \cite{nakamoto2008bitcoin}, the world's most successful
cryptocurrency, is a popular payment method and is currently processing on average more than 300k transactions per day.
Payment Processors such as BitPay and Coinbase offer online store integration on platforms such as Shopify, OpenCart, and WordPress eCommerce for sending and receiving bitcoins.
This service is popular amongst merchants willing to accept cryptocurrencies as a form of payment as it automatically converts bitcoins to fiat currency and removes the risks involved in Bitcoin's price volatility.
The total Bitcoin transaction volume of such payment processors has reportedly been worth around \$10M per day on average in 2019~\cite{chainalysis20}.

Major Payment Processors that mediate the service between user wallets and merchants require user wallets to implement the BIP21 URI Scheme~\cite{bip21:payment} and BIP70 Payment Protocol~\cite{bip70:payment} Bitcoin community standards.

BIP21 provides a two-step payment procedure in which the user follows a link that triggers the user's wallet to automatically prepare a payment of the correct amount to the correct address, both embedded within the URI.
BIP70 goes further and enables authenticated communication with the merchant and improved payment experience including receipt notifications, refund addresses, and payment acknowledgement.
BIP70 hence improves on BIP21 in two major areas: security and usability.
Since BIP21 does not provide any form of authentication, it is open to man-in-the-middle attacks.
Malicious third-party scripts and extensions, viruses, or malicious Tor exit nodes have been reported to mount such attacks and change the receiver's Bitcoin
address to route funds towards adversaries~\cite{malicious14extension,virus16bitpay}.
BIP70 however, builds in the authentication of the merchant using X.509 certificates.
Besides, it improves on the usability of Bitcoin payments as customers are no longer required to manually handle Bitcoin addresses, consisting of 26--35 random-looking alphanumeric characters, used to send and receive bitcoins.
Instead, the customer can verify the merchant's identity using a human-readable name, coming from the certificate's `common name' field, before authorising a payment.
A refund Bitcoin address is also sent to the merchant by the user's wallet that should be used in the event of a refund.
Indeed, BitPay has reported a sharp reduction of payment errors, including under- and over-payments, as a result of the adoption of the BIP70 Payment Protocol~\cite{bitpay18payment}.
BIP70 is hence supported by major payment processors such as BitPay and Coinbase, along with popular Bitcoin development libraries such as BitcoinJ. 

In a recent setback in the widespread adoption of BIP70, support for it was removed from the Bitcoin Core client, the Bitcoin reference implementation, in version 0.20.0 released in June 2020~\cite{bitcoin20core}.
However, the continued use of the protocol is still supported by BitPay and Coinbase wallets, many other popular (software) wallets (see e.g.~\cite{pp-wallets-bitpay}), and major hardware wallets such as Trezor.

From a high-level point of view, the Payment Protocol essentially provides two pieces of evidence that can be used by the parties involved to prove they have followed the protocol without malice: the \emph{Payment Request} sent by the merchant to the customer which is digitally signed by the merchant, and the payment transaction broadcast by the customer and included in the blockchain which is digitally signed by the (pseudonymous) customer.
In our previous work ~\cite{mccorry2016}, we demonstrated that a third piece of evidence is further required: endorsement of the refund address(es) by the customer. Without this third piece of evidence two types of attacks would be possible:
\begin{itemize}
\item An attack that allows a customer to request refunds to an illicit trader's Bitcoin address for a previous payment, e.g. by cancelling their order. This is called the \textit{Silkroad Trader} attack and leverages the fact that the customer can later deny providing an illicit trader's address for refund purposes.
\item An attack that allows a rogue trader to forward a \emph{Payment Request} from an honest merchant to a customer and later request refunds from the merchant to their own Bitcoin address. This is called the \textit{Marketplace Trader} attack and leverages the fact that refund requests are not authenticated.
\end{itemize}

Responding to the disclosure of the above attacks, payment processors tightened their refund address communication policy and to some extent mitigated the Marketplace Trader attack.
The Silkroad Trader attack however, is still an open issue and addressing it would require a revision of the Payment Protocol.

We focus on the Silkroad Trader attack and extend our previous work \cite{mccorry2016} to provide the first formal model of BIP70. The Payment Protocol security analysis, which is performed using the symbolic model-checker OFMC, confirms the above attack as an Authentication Attack. Similarly, we also verify that the revised Payment Protocol in \cite{mccorry2016} fixes the vulnerability. This is the first formal model for an application of the Bitcoin blockchain and is complementary to other work in the research community which includes formal models for Nakamoto-style consensus protocols \cite{garay2015bitcoin}, formal verification of the runtime and functional correctness \cite{bhargavan2016formal,Hildenbrandt2018,Tsankov2018,luu2016making} for smart contracts in other cryptocurrencies due to substantial thefts, and formal languages for writing Bitcoin script~\cite{OConnor2017}.

\paragraph{Contributions} This work builds on and extends our previous work (McCorry et al.~\cite{mccorry2016}) focusing on the \textit{Silkroad Trader} attack.
The contributions presented are summarised below:
\begin{itemize}
\item We present the first formal model of the Bitcoin Payment Protocol. It utilises
communication channels as abstractions conveying security goals that allow us
 to specify a model that is tractable and can be analysed more efficiently by the model-checker OFMC.
\item We demonstrate, by model-checking, that the protocol is vulnerable to authentication attacks of the refund addresses.
This attack was informally presented in \cite{mccorry2016} as the \textit{Silkroad Trader} attack.
\item We validate the revised Payment Protocol proposed in \cite{mccorry2016} and confirm that it prevents the \textit{Silkroad Trader} attack.  Moreover, we propose a simpler alternative fix, and the model-checking shows that no further attacks to the identified security goals can be performed once any of the two fixes is applied.
\end{itemize}

\section{Background\label{sec:Background}}
We briefly discuss background information about Bitcoin, the formal modelling
technique used in this paper, and related work in these areas before presenting the Payment Protocol standard in the next section.

\subsection{Bitcoin}
We briefly introduce three key concepts:
\emph{Bitcoin addresses}, a form of pseudonymous identification,
\emph{transactions}, a mechanism used to record the transfer of bitcoins, and the
\emph{blockchain}, a decentralised data structure storing all transactions on the network.

A \emph{Bitcoin address} is an identifier in the Bitcoin network and is computed from the hash of an Elliptic Curve (EC) public key.
An address serves as a pseudonymous identified of the user in possession of the corresponding private key.
The corresponding private key can be used to claim bitcoins sent to a user and to authorise payments to other users using the Elliptic Curve Digital Signature Algorithm (ECDSA).
Given the output length of the hash functions involved in the computation, the probability of collision is negligible, so these identifiers can be safely assumed unique within the network.

\begin{figure}[t]
\begin{centering}
\includegraphics[width=0.7\textwidth]{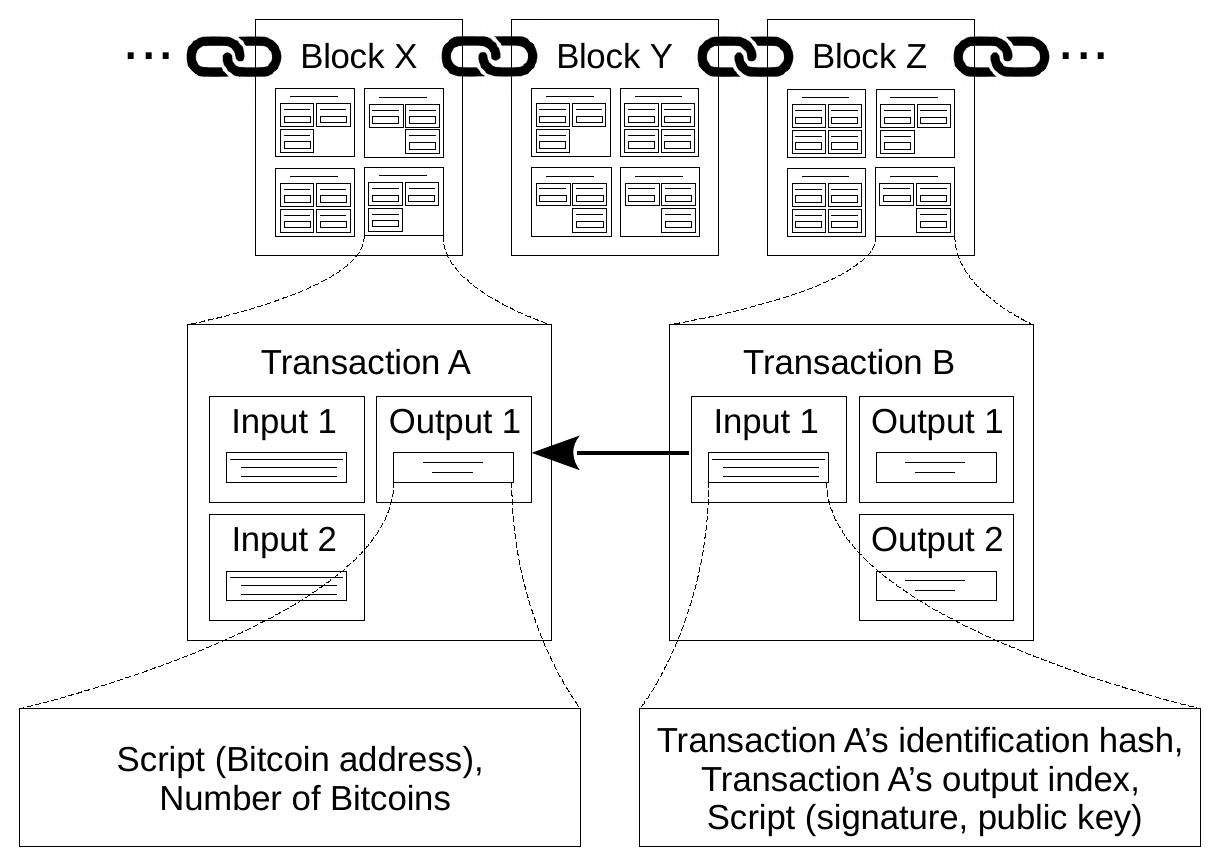}
\par\end{centering}
\caption{\label{fig:transactions}Information stored in the inputs and outputs
of Bitcoin transactions}
\vspace{-\baselineskip}
\end{figure}

A \emph{transaction} records the transfer of bitcoins.
It consists of one or more inputs, specifying the source of bitcoins being spent, and one or more outputs, specifying new owner's Bitcoin address and the amount being transferred (see Figure~\ref{fig:transactions}).
To authorise the payment, the sender must specify an input consisting of the previous transaction's identification hash and an index to one of its outputs, and provide the corresponding public key and a valid digital signature.
The inputs and outputs are controlled by means of scripts in a Forth-like language specifying the conditions sufficient to claim the bitcoins.
The standard script is the \textsf{pay-to-pubkey-hash} requiring a single signature from a Bitcoin address to authorise the payment.

The \emph{blockchain} stores the complete transaction history of the network with a secure time-stamp~\cite{nakamoto2008bitcoin} arranged in blocks of transactions.
This append-only data structure (\emph{ledger}) is stored in a distributed fashion by most users of the network.
Appending new transactions requires solving a \emph{proof of work} puzzle which is computationally difficult to solve but easy to verify if a solution is given.
Nodes that solve proofs of work are called \emph{miners}.
They receive rewards in Bitcoin for their computational effort.

\subsection{Formal Modelling Approach}
\label{sec:formal_modelling}
Our approach to formal modelling and security analysis of the BIP70 Payment Protocol involves
the symbolic model-checker OFMC \cite{basin2005ofmc} (version 2020),
and the specification of a model in \emph{AnB} \cite{anb}, a formal language
in the style of \emph{Alice and Bob} narrations. An important reason
for adopting this methodology is that in the specification of the protocol,
it is possible to model communication channels as abstractions conveying
security goals like authenticity and/or secrecy, without the need
to specify the concrete implementation used to enforce such goals.
The protocol modeller and the verification tool can then rely on the
assumptions provided by such channels. This allows to specify a simpler model
that is tractable by the verifier, and can be analysed more efficiently.
In fact, modelling the underlying channel cryptographic implementation explicitly will
lead model the checker towards a state-explosion problem and/or face out-of-memory errors.

Another important feature is that in AnB channels, agents can
be identified by pseudonyms rather than by their real identities, similarly
to what happens in secure channels like TLS without client authentication.
Such ability to model and verify a range of different channels makes
AnB suitable for the verification of payment protocols like BIP70,
since in such protocols secure channels (HTTP over TLS) are used
and agents use pseudonyms (e.g. ephemeral public keys).

Moreover, since we model BIP70 protocol on top of these channels,
we need to discuss whether the vertical composition of such protocols
is secure.

\paragraph{Channel as assumption}

In general, OFMC allows specifying three type of channels in AnB: \emph{authentic},
\emph{confidential}, and \emph{secure}, with variants that allow agents
to be identified by a pseudonym rather than by a real identity. The
supported standard channels are:
\begin{enumerate}
\item $A\insecCh B:M$, an insecure channel from $A$ to $B$, under the
complete control of a Dolev-Yao intruder \cite{dolev83ieee};
\item $A\athCh B:M$, an authentic channel from $A$ to $B$, where $B$
can rely on the fact that $A$ has sent the message $M$ and meant
it for $B$;
\item $A\cnfCh B:M$, a confidential channel\emph{, }where $A$ can rely
on the fact that only $B$ can receive the message \emph{M};
\item $A\secCh B:M$, a secure channel (both authentic and confidential).
\end{enumerate}
Pseudonymous channels \cite{moedersheim2009secure} are similar to
standard channels, with the exception that one of the secured endpoints is logically tied
to a pseudonym instead of a real name. The notation $[A]_{\psi}$
represents that an agent $A$ is not identified by its real name $A$
but by the pseudonym $\psi$. Usually $\psi$ can be omitted, simplifying the notation to
$[A]$, when the role uses only one pseudonym for the entire session,
as it is in our case and in many other protocols.

For example, $[A]\athCh B:M_{1}$ denotes an authentic channel from
$A$ to $B$, where $B$ can rely on the fact the an agent identified
by a pseudonym has sent a message $M_{1}$ and this message was meant
for B. If during the same protocol run, another action like $[A]\athCh B:M_{2}$
is executed, $B$ can rely on the fact that the same agent (identified
by the same pseudonym) has also sent $M_{2}$, and again the message
was meant for \emph{B}.

Assuming that \emph{$B$} does not know already the real name of \emph{A},
the execution of these two actions does not allow\emph{ $B$} to learn
the real identity of $A$ (unless this information is made available
during the protocol execution), but $B$ has a guarantee that he
was communicating with the same agent during both message exchanges.
The term \emph{sender invariance} is used to refer to this property,
and the most common example is the TLS protocol without client authentication.

\paragraph{Vertical Protocol Composition}

Since in our model BIP70 runs on top of abstract channels providing
security guarantees (such as the TLS without client authentication),
we are in effect vertically composing TLS and BIP70 protocols.
Strictly speaking, we should consider HTTPS, but since the security
guarantees are provided by TLS, the model can simply abstract the
HTTP messages.

In general, given a secure protocol $P_{1}$ that provides a certain
channel type as a goal and another secure protocol $P_{2}$ that
assumes this channel type, their vertical composition $P_{2}[P_{1}]$
is not secure as attacks may be possible even when the individual
protocols are all secure in isolation. Sufficient conditions for vertical
composition have been established~\cite{Moedersheim2014} and, in
essence, they require the disjointness of the message formats of $P_{1}$
and $P_{2}$, and that the payloads of $P_{2}$ are embedded into
$P_{1}$ under a unique context to define a sharp borderline. According
to M{\"o}dersheim and Vigan{\`o}~\cite{Moedersheim2014}, these conditions and the other minor conditions
are satisfied in practice by a large class of protocols. As the specific
implementation of the underlining protocol is not part of the BIP70
specification ($P_{2}$), but $P_{2}$ only assumes that the communication
occurs on channels that guarantee a secret communication with server
authentication ($P_{1}$), we make our analysis under the assumption
that the conditions sufficient for vertical composition specified in~\cite{Moedersheim2014}
are satisfied.

\subsection{Related Work\label{sec:RelatedWork}}

An overview of related research on Bitcoin payment protocols and formal methods applied to Bitcoin and blockchain technologies is given below.

\subsubsection{Bitcoin Payment Protocols\label{sec:BitcoinRelatedWork}}

The Payment Protocol is designed for `on-chain' payments in which all the transactions required for the intended payment are appended to the blockchain.
Inherent and practical limitations on global transaction rates translate into serious scalability issues  for Bitcoin and other cryptocurrencies. This has served as the main motivation for a line of work on `off-chain' payment channels (See~\cite{McCorryMSH16PayNet} for an overview), in which payments are optimistically carried out with limited interaction with the blockchain and `on-chain' transactions are only used to resolve party failures or to settle disputes.
Recent proposals in this area include AMCU~\cite{amcu} for Bitcoin (and other cryptocurrencies with restricted scripting capabilities) and Sprites~\cite{miller2019sprites} for Ethereum.
The most widely deployed of such networks are Lightning~\cite{lightning16} for Bitcoin and Raiden (see \url{raiden.network}) for Ethereum.
Although these alternative payment methods have a growing user base, their overall usage remains comparatively low (see e.g. statistics in~\cite{lightning-usage-stats-cryptonews,lightning-usage-stats-bitmex}).

Lack of methods for post-payment communications that are securely bound to the original payment transaction has been acknowledged by the community.
In~\cite{McCorrySCH15AKE}, the authors propose to bootstrap authenticated key exchange protocols between the sender and the receiver of an existing transaction leveraging the signatures recorded on the blockchain.
Such a protocol will provide a secure channel between the parties to a transaction and can be used for secure post-transaction communications including arranging refunds.
However, such protocols have not been deployed in practice.

An early solution to the lack of refund address endorsement by Hearn~\cite{overkill} suggested endorsement by any key that authorised the original payment transaction.
However, this solution was shown to be prone to the `malicious co-signer attack'~\cite{mccorry2016}.
Subsequent to~\cite{mccorry2016}, where the vulnerability of the Payment Protocol to refund attacks was demonstrated, alternative mitigation methods have been also proposed. In~\cite{AvizhehSS18}, Avizheh et al.\ proposed another solution based on multi-signature and time-locked transactions.
The idea is that in case of a refund request, the merchant prepares two refund transactions: one that requires signatures from both the refundee and the (original) customer to be claimed (a `multi-signature' transaction), and another that enables the customer to claim the refund after certain period of time (a `time-locked' transaction) in case the customer does not authorise the former refund transaction.
This would reduce the amount of log keeping the merchant needs to implement to protect itself against refund attacks compared to the solutions we proposed in \cite{mccorry2016}.
However, Avizheh et al.'s solution would need substantial changes to the Payment Protocol standard, whereas our modifications are designed to require minimal changes to the standard.

Another noteworthy related service is the Ethereum Name Service (see \url{ens.domains}). This service provides a secure binding between the domain name of a merchant and their cryptocurrency addresses, supported by Ethereum smart contracts. Such a service is, in effect, akin to a distributed DNS service. A customer only needs to input the domain name of a merchant as the recipient of a payment in a wallet that supports the Ethereum Name Service look-up protocol. The wallet would be able to find the corresponding authenticated address via communication with the Ethereum blockchain. Although such services address the low usability and lack of authentication of \emph{merchant} Bitcoin address, they do not offer any solution for customer refund address authentication.

In summary, while the Payment Protocol remains the dominant Bitcoin payment method, especially with major payment processors, \emph{customer refund address authentication} remains an open problem in practice. Existing proposals for secure post-payment authentication are not deployed and alternative refund address authentication mechanisms require substantial changes to the established standard. Our proposed modifications to the standard are minimal and can be readily adopted. Furthermore, all previous works addressing refund address authentication have followed the `design-break-fix' paradigm in which solutions merely guarantee that specific attack strategies do not apply anymore. We break from this paradigm and provide a formal modelling of the required security properties along with a verification of our proposed solutions satisfying those formalised properties. Such verification provides a guarantee of security \emph{independent of the attack strategy} for a general class of adversaries specified by their capabilities and goals, namely the Dolev-Yao attacker model~\cite{dolev83ieee}. Interestingly, the model is also realistic. In fact, Herzog~\cite{Herzog2005} proved that there are many significant cases in which the Dolev-Yao adversary can be a valid abstraction of all realistic adversaries.

\subsubsection{Formal Modelling and Verification\label{sec:FormalRelatedWork}}
Given the growing interest in blockchain and cryptocurrencies, these technologies have been the subject of studies by the formal methods community as well, Bitcoin in particular.
These works either focus on Bitcoin transactions, blockchain, and their security properties, or consider other components of the cyrptocurrency ecosystem such as consensus mechanisms, smart contracts, and wallets. 
None of these works attempt formalising any of the Bitcoin protocols built \emph{on top of the core Bitcoin protocol} such as the Payment Protocol. 
However, we briefly review these works in the interest of completeness. 

Among the previous works considering the formalisation of the core Bitcoin ecosystem are the following. 
Garay et al.~\cite{garay2015bitcoin} presented a formal modelling of the Bitcoin backbone,
the protocol used at the core of Bitcoin's transaction ledger. They formalised and
proved basic properties they called `common prefix', `chain quality', and `chain growth',
analyzing applications that can be built on top of the backbone protocol, focusing on Byzantine agreement (BA)
and on the notion of a public transaction.
Atzei et al.~\cite{Atzei2018} proposed a formal model for Bitcoin transactions that abstractly describes
their essential aspects, and at the same time enables formal reasoning. The model allows formally proving several well-formedness properties of
the Bitcoin blockchain, for instance that each transaction can only be spent once.
Chaudhary et al.~\cite{Chaudhary2015} also considered the success probability of a double spending attack, which is linked
to the computational power of the attacker. As the validation of Bitcoin transactions requires the successful execution of scripts,
Klomp and Bracciali~\cite{Klomp2018} worked on the formal verification for the Bitcoin validation framework, proposing
a symbolic verification theory and a toolkit for the verification.

Other researchers have investigated different components of the cyrptocurrency ecosystem. For example, Duan and al.~\cite{duan2018formal} presented the model-based formalization, simulation and verification of a blockchain protocol by using the SDL formalism of Telelogic/Rational Tau considering aspects such as security and safety of blockchain. The work also provides support for assessing different network consensus algorithms as well as on the topology of blockchain networks. Additionally, a number of works focused on formal modelling of wallets \cite{Turuani2016,Arapinis2019} and smart contracts \cite{Harz2018,dwivedi2019formal}.

In summary, no previous work has considered formal modelling and verification of protocols built on top of the core Bitcoin protocol such as the Payment Protocol. Our previous work~\cite{mccorry2016} has shown that even if the core Bitcoin protocol is considered secure, vulnerabilities can exist in the design of the Payment Protocol which is built on top of the core Bitcoin protocol. It remained an open problem to formally model the amended Payment Protocol and verify that it satisfies the intended security properties assuming that the core Bitcoin protocol is secure. 
In this paper, we address this gap.

\section{The Payment Protocol\label{sec:Payment-Protocol}}
The Payment Protocol was proposed in 2013 by Andresen and Hearn in BIP70~\cite{bip70:payment} and later adopted by the Bitcoin community as a standard.
The authors present the goal of the protocol as follows:
\begin{quote}
\emph{``This BIP describes a protocol for communication between a
merchant and their customer, enabling both a better customer experience
and better security against man-in-the-middle attacks on the payment
process.'' }
\end{quote}
The communication channel between the customer and merchant is strongly recommended to be over HTTPS leveraging the merchant's X.509 certificate issued by a trusted Certificate Authority. This allows the customer to authenticate messages from the merchant.

\begin{figure}[t]
\centering
\includegraphics[width=0.9\textwidth]{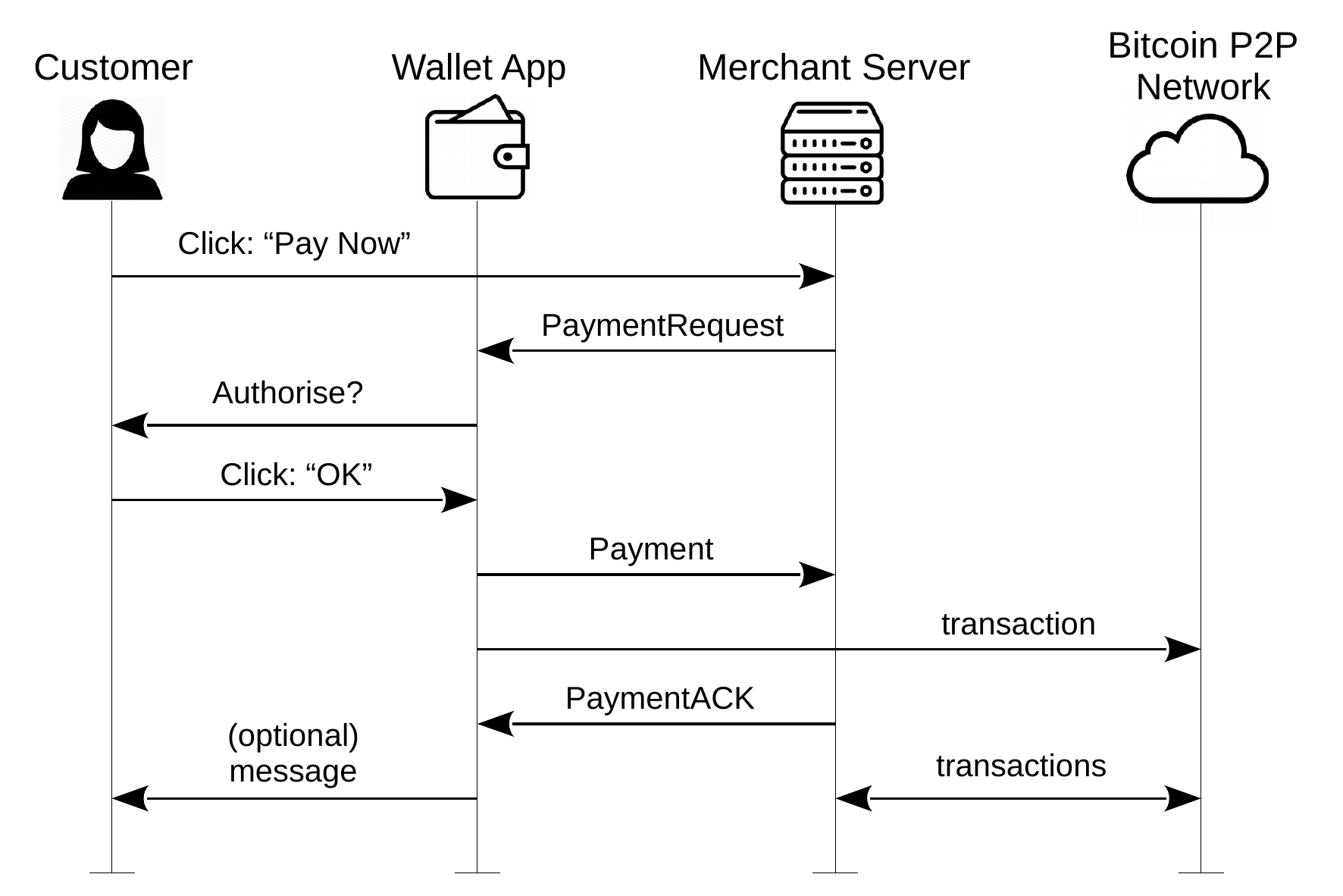}
\caption{\label{fig:paymentprotocol}Overview of the Payment Protocol (adapted from~\cite{bip70:payment})}
\vspace{-\baselineskip}
\end{figure}

Figure \ref{fig:paymentprotocol} outlines an overview of the messages exchanged and actions performed during the protocol execution.
The protocols begins with the customer clicking on the `Pay Now' button on the merchant's website to generate a Bitcoin payment URI.
This URI allows to open the customer's Bitcoin wallet and download the \textit{Payment Request} from the
merchant's website.
The wallet app can then verify the digital signature on the \textit{Payment Request} using the public key of the merchant and the validity of the associated certificate.
Given successful verification of the signature, the merchant's name in a human-readable format, extracted from the X.509 certificate's `common name' field, and the bitcoin amount requested are shown to the customer requesting for authorisation of the payment.
When the user authorises the payment, the wallet puts together a payment transaction and broadcasts it to the network.
Besides, it includes the transaction and refund addresses within a \textit{Payment} message which is sent back to the merchant.
The merchant then replies to the customer wallet with a \textit{Payment Acknowledgement} and once the payment transaction is detected on the blockchain the customer receives a confirmation of the payment.

\subsection{Modelling BIP70}\label{subsec:ModellingBIP70}
\begin{table}[t]
\centering
\begin{longtable}{l>{\raggedright}p{8.5cm}}
\hline
Identifier & Description\tabularnewline[\doublerulesep]
\hline
$\bitcoinAddress{}_{\bob}$ & merchant Bitcoin address for the current transaction,
a public key freshly generated by $\bob$ with the corresponding private key denoted by $\inv{\bitcoinAddress_{\bob}}$ \tabularnewline[\doublerulesep]
$\bitcoinAddress{}_{\alice_{i}}$ & customer $\alice_{i}$ Bitcoin address for the current transaction,
a public key freshly generated by $\alice_{i}$, with
the corresponding private key denoted by $\inv{\bitcoinAddress_{\alice_{i}}}$\tabularnewline[\doublerulesep]
$\refundAddress_{\alice_{i}}$ & refund address of customer $\alice_{i}$\tabularnewline[\doublerulesep]
$\bitcoinB$ & number of Bitcoins for the current transaction\tabularnewline[\doublerulesep]
$\bitcoinB_{\alice_{i}}$ & number of Bitcoins to be refunded to $\refundAddress_{\alice_{i}}$ in case of a refund\tabularnewline[\doublerulesep]
$t_{1}$, $t_{2}$* & timestamps indicating \emph{Payment Request} creation and expiry times, resp.\tabularnewline[\doublerulesep]
$m_{\bob}$* $m_{\alice}$*, $m'_{\bob}$* & memo messages included in the \emph{Payment Request} (by $\bob$), \emph{Payment} (by $\alice$), and \emph{Payment Acknowledgement} (by $\bob$) messages\tabularnewline[\doublerulesep]
$u_{\bob}$* & payment URL\tabularnewline[\doublerulesep]
$z_{\bob}$* & payment id provided by the merchant\tabularnewline[\doublerulesep]
\hline
\end{longtable}
\caption{Notation -- Identifiers used to denote the data exchanged (optional parameters are starred)}
\label{tab:notation}
\end{table}
Our general approach to formal modelling, similarly to~\cite{AnBEMV}, requires the analysis
of the protocol specification and its informal security requirements.
In particular, in order to verify the BIP70 protocol, we build a model with \emph{$n$
}$\left(n\geq1\right)$ customers $C_{1},\ldots,C_{n}$ and one merchant
$M$. We assume that these agents can trade over Bitcoin and that
the identity of the merchant is known to all of them. Strictly speaking,
since multiple customers might be co-operating in the payment with
a single merchant, our model requires that at least one of customers
knows the merchant name. We also assume that, prior the run of the
protocol, the merchant does not know the identity of these customers
and the communication between agents and merchants does not require
a mechanism which explicitly discloses the real identity of the client.
Such one-way authentication can be customarily achieved using HTTPS, as in BIP70.
In this case, the client is guaranteed that messages
are exchanged with the authenticated server, but the server is only
guaranteed that the communication channel is shared with the same
pseudonymous agent. In our case, we denote the pseudonym
of the agent $C_{1}$ during the protocol run by $[C_{1}]$.

In the model, we also assume that $C_{1}$ is the only agent that communicates
with the merchant, while other agents communicate with $C_{1}$ using
a secure channel (or out-of band) to collaboratively setup an order
for the merchant.
This reflects the actual usage scenario of the Payment Protocol in which payment may be made from multiple psudonymous Bitcoin addresses, belonging to one or multiple actual entities, and it is the responsibility of the customer communicating with the merchant to assemble the payment transaction in coordination with all the Bitcoin address holders.
The model employs two kinds of channels:
\begin{itemize}
\item $[C_{1}]\secCh M$ denotes a secure (secret and authentic) channel between
the client $C_{1}$ and the merchant $M$, with the peculiarity that \emph{$M$}
can bind the other end point to a pseudonym \emph{$[C_{1}]$} rather
than to the real identity of $C$.
\item $C_{i}\secCh C_{j}$ denotes a secure channel between the clients \emph{$C_{i}$}
and \emph{$C_{j}$}.
\end{itemize}
We use the identifiers listed in Table~\ref{tab:notation} to denote the data exchanged.

Moreover, we denote the hash function used in generating Bitcoin addresses by $\hashsymbol$.
Let us introduce the following definitions used in the protocol
specification:
\begin{itemize}
\item
  $\omega_{i}=\bitcoinB_{\alice_{i}},\hashsymbol(\bitcoinAddress{}_{\alice_{i}})$: the previous transaction outputs for customer $\alice_{i}$;
\item
  $\transaction_{\alice_{i}}=\trNA\left(\omega_{i}\right)$:
  the previous transaction for customer $\alice_{i}$.
  Future transactions depend only on unspent/spendable transaction outputs;
  we consider here a function $\trNA$ that returns a transaction parameterised on the output used by $\alice_{i}$ in the current transaction;
\item
  $\inputTransaction_{\alice_{i}}=\sign{\inv{\bitcoinAddress_{\alice_{i}}}}{\hashsymbol(\transactionid_{\alice_{i}}),{\bitcoinAddress{}_{\alice_{i}}}}$:
  the transaction input endorsed by $\alice_{i}$;
\item
  $\pi=\inputTransaction_{\alice_{1}},...,\inputTransaction_{\alice_{n}}$:
  the transaction input, a list transaction inputs endorsed by the customers;
\item
  $\payreq=\sign{\inv{\sk{\bob}}}{\hashsymbol(\bitcoinAddress_{\bob}),\bitcoinB,t_{1},t_{2},m_{\bob},u_{\bob},z_{\bob}}$:
  the \emph{Payment Request}, a message digitally signed with $\inv{\sk{\bob}}$,
  the private key of $\bob$.  The associated public key utilised to verify the digital signature,
  that we denote as $\sk{\bob}$, is certified by a Certificate Authority and stored in a X.509 certificate;
\item
  $\refundAuth{}_{\alice_{i}}=(\refundAddress_{\alice_{i}},\bitcoinB_{\alice i})$:
  the refund address and amount for customer $\alice_{i}$;
\item
  $\transaction_{\alice}=\pi,(\hashsymbol(\bitcoinAddress_{\bob}),\bitcoinB)$:
  one or more valid transactions, where $\pi$ represents the inputs, and $(\hashsymbol(\bitcoinAddress_{\bob}),\bitcoinB)$ represent the output.
\end{itemize}

\subsubsection{Agents' Initial Knowledge}

The initial knowledge of a model with one merchant $\bob$ and two
customers $\alice_{1},\alice_{2}$ is as follows:
\begin{itemize}
\item $\alice_{1}:\alice_{1},\alice_{2},\bob,\hashsymbol,\trNA,\skNA,\paynow$
\item $\alice_{2}:\alice_{1},\alice_{2},\hashsymbol,\trNA,\skNA$
\item $\bob:\bob,\hashsymbol,\trNA,\skNA,\inv{\sk{\bob}},\paynow,t_{1},t_{2}$
\end{itemize}
Each agent has an identity and access to the hash function $\hashsymbol$,
the symbolic function $\trNA$ and a symbolic function $\skNA$ for
modelling digital signatures.%

{} In particular, the $\skNA$ function allows customers $\alice_{i}$
to retrieve $\sk{\bob}$ the public key of agent $\bob$ from a repository,
and verify the corresponding X.509 certificate, provided that they
know the name of $\bob$.

$\inv{\sk{\bob}}$ represents the private key of $\bob$ and is
known only by $\bob$. It should be noted that in the \emph{AnB} language,
$\inv$ is a private function. Therefore, neither other agents nor the intruder can
use $\invNA$ to retrieve any agent's private key.

Initially, $\bob$ does not know the identities $\alice_{1}$ and $\alice_{2}$,
while $\alice_{1}$ and $\alice_{2}$ know each other as they need
to collaborate to build the transaction. However, only $\alice_{1}$
knows $\alice_{2}$ since $\alice_{1}$ will be the only customer
interacting with the merchant. Finally, various constants ($t_{1},t_{2},\paynow$)
are available to agents according to the protocol specification.

The initial knowledge can be easily generalised for $n$ customers;
it should be noted that a customer does not need to know all other
customers prior the protocol run, but at least one. As customers can
coordinate as they wish (including out-of-band communication), only one customer
will need to interact with the merchant.

\subsubsection{Security Goals\label{subsec:Security-Goals}}

We expect the following security goals to hold after the protocol execution:

 \begin{itemize}
\item \textbf{Goal 1:} \emph{Refund Addresses Authentication}. $\bob$ has a guarantee that all refund addresses $\refundAddress_{\alice_{i}}$,
for all $i=1..n$ are provided by and linked to the customers involved
in the transaction;
In \emph{AnB}, we denote the goal as:\\
\vskip -3mm
$\bob$ \texttt{weakly authenticates} $\alice_{i}$ \texttt{on} $\refundAddress_{\alice_{i}},\bitcoinB_{\alice_{i}}$
(for all $i=1..n$ )
\vskip 1mm
\item \textbf{Goal 2:} \emph{Refund Address Agreement and Secrecy}. All refund addresses $\refundAddress_{\alice_{i}}$ are secret and
known only by the merchant and the customers involved in the transaction. In \emph{AnB}, we denote the goal as:\\
\vskip -3mm
$\left(\refundAddress_{\alice_{1}}, \ldots ,\refundAddress_{\alice_{n}}\right)$
\texttt{secret between} $\bob,\alice_{1}, \ldots ,\alice_{n}$\\
\end{itemize}

Note that the Payment Protocol is built on top of the core Bitcoin protocol and blockchain and the question we consider is whether the Payment Protocol is secure assuming the core Bitcoin protocol is secure. 
Therefore, we do not model the security goals that are expected to be guaranteed by the core Bitcoin protocol and blockchain, such as the double-spending prevention, and assume that they are satisfied. In fact, as discussed earlier, the security properties of the core Bitcoin protocol and blockchain have been formally proven in previous works~\cite{Atzei2018,Chaudhary2015}.  
By the same token, we do not explicitly consider the security issues at the lower layers of the networking stack such as eavesdropping, prediction and fixation, since our work only concerns the application layer and assumes that protocols such as TLS are secure. 
The approach of considering the security properties of different layers in isolation is sound, provided that the conditions of the vertical composition theorem~\cite{Moedersheim2014} are satisfied as discussed in Section~\ref{sec:formal_modelling}.
It should be noted that the secrecy goal (2) prevents eavesdropping, and that known prediction and fixation vulnerabilities have been addressed by more recent versions of TLS~\cite{Bhargavan2017,Cremers2017}.

\subsubsection{Protocol Actions}

Given their initial knowledge, agents are involved in a sequence of
message exchanges over the designated channel. On the sender side,
agents should have enough information to compose the message. On the
recipient side, every agent must decompose the incoming messages (for
example, decrypting the message or verifying a digital signature)
according to their current knowledge, including knowledge acquired
during previous steps. For simplicity, we assume that all public
keys are available, at a certain point of the protocol execution,
to the agents and the intruder.

\medskip{}

\begin{tabular}{llr}
$[\alice_{1}]\secCh\bob$ & :\ $\paynow$ & \emph{$\alice_{1}$ clicks }`Pay Now'\tabularnewline[10pt]
$\bob\secCh[\alice_{1}]$ & :\ $\payreq$ & \emph{Payment Request} \tabularnewline[10pt]
$\alice_{1}\secCh\alice_{2}$ & :\ $\refundAddress_{\alice_{1}}\bob,\payreq,\bitcoinAddress{}_{\alice_{1}}$ & $\alice_{1},\alice_{2}$ \emph{cooperate -}\tabularnewline[10pt]
$\alice_{2}\secCh\alice_{1}$ & :\ $\refundAddress_{\alice_{2}},\inputTransaction_{\alice_{2}}$ & \emph{- to build a transaction}\tabularnewline[10pt]
$[\alice_{1}]\secCh\bob$ & :\ $z_{\bob},\transaction_{\alice},\refundAuth_{\alice_{1}},\refundAuth_{\alice_{2}},m_{\alice}$ & \emph{Payment} \tabularnewline[10pt]
$\bob\secCh[\alice_{1}]$ & :\ $z_{\bob},\transaction_{\alice},\refundAuth_{\alice_{1}},\refundAuth_{\alice_{2}},m_{\alice},m'_{\bob}$ & \emph{PaymentACK} \tabularnewline[10pt]
\end{tabular}

\medskip{}

\begin{figure*}[t]
\centering
\includegraphics[scale=0.9]{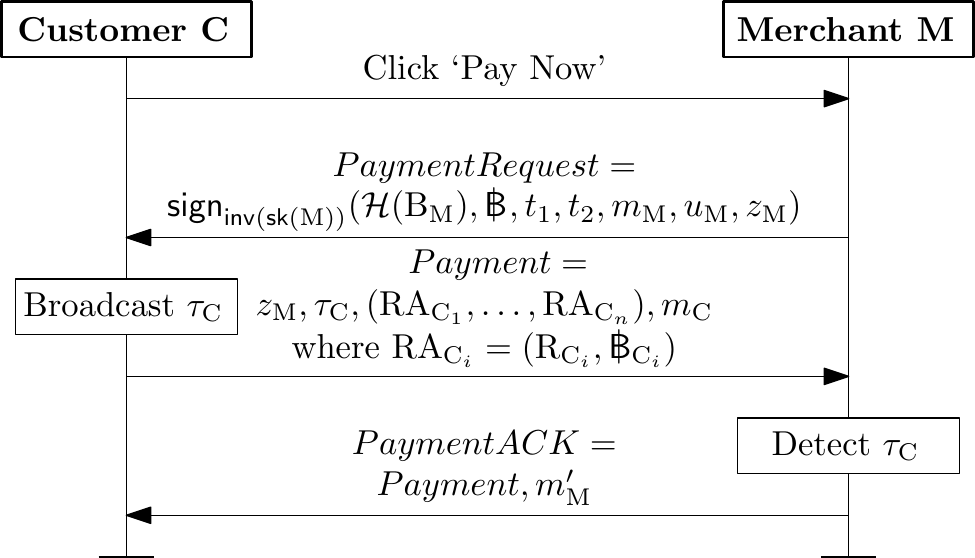}
\caption{
    \label{fig:messages}%
    Expanded message contents for the Payment Protocol for $\alice$ and $\bob$. Messages are sent over an HTTPS communication channel.
    We use the notation $\sign{\inv{\sk{\bob}}}{X}$ to denote both the message $X$ and the digital signature on the message by the private key $\inv{\sk{\bob}}$.
    }
\end{figure*}

\subsubsection{Protocol Message Details}
The Payment Protocol standard specifies the format of the \emph{Payment Request}, \emph{Payment}, and \emph{Payment Acknowledgement} messages.
The standard only \emph{recommends} running the protocol over HTTPS, however in this paper we assume this is always the case.
Discussed attacks apply regardless of whether HTTPS is used.
Although the standard supports payment via multiple transactions, we discuss the details of the messages here for the case where the customers pays through only one transaction.
The proposed solutions and formalisation results can be easily extended to the case where a payment is made through multiple transactions.
The protocol messages are as follows:
\begin{itemize}
\item The \emph{Payment Request} consists of the recipient's Bitcoin address $\hashsymbol(\bitcoinAddress_{\bob})$, requested Bitcoin amount \bitcoinB, timestamps $t_{1}, t_{2}$ corresponding to the request creation and expiry times, a `memo' field $m_{\bob}$ for a note showed to the customer, the payment URL $u_{\bob}$ where the payment message should be sent, and an identifier $z_{\bob}$ for the merchant to link subsequent payment messages with this request.
All of the fields are collectively signed by the merchant using their private key denoted by $\inv{\sk{\bob}}$ corresponding to their X.509 certificate public key.

\item The \emph{Payment} message consists of the merchant identifier $z_{\bob}$, the payment transaction $\transaction_{\alice}$, a list of pairs of the form $\refundAuth{}_{\alice_{i}}=(\refundAddress_{\alice_{i}},\bitcoinB_{\alice i})$ each containing the refund address $\refundAddress_{\alice_{i}}$ and the amount to be paid to that address $\bitcoinB_{\alice i}$ in case of refund, and an optional customer `memo' field $m_{\alice}$.

\item The \emph{Payment Acknowledgement} consists of a copy of the \emph{Payment} message sent by the customer and an optional `memo' $m'_{\bob}$ to be shown to the customer.
\end{itemize}

The Payment Protocol messages are shown Figure~\ref{fig:messages}.
Note that the \emph{Payment} message, and specially the refund addresses provided therein, are not signed by the customer, and although protected by HTTPS, they can be subsequently repudiated by the customer.
This is the underlying weakness that allows the Silkroad Trader attack.

\section{Verification of the Silkroad Trader Attack\label{sec:attackpayment}}
We now discuss the Silkroad Trader attack, proposed in \cite{mccorry2016} and verify, by model-checking, the violation of the security goals.
The Silkroad Trader attack allows a customer to route Bitcoin payments through an honest merchant to an illicit trader and later deny their involvement as we discuss below.
We also demonstrate that this attack can be captured as an authentication attack within our formal model of the Payment Protocol.

\subsection{Silkroad Trader Attack\label{subsec:Silkroad-Trader-Attack}}
As mentioned earlier, the refund addresses provided by the customer are not digitally signed.
This means that a malicious customer will be able to order refunds to any arbitrary address without being required to provide any undeniable authorisation.
The Silkroad Trader attack leverages such plausible deniability afforded to the customer in the Payment Protocol.

\begin{figure*}[t]
\centering
\includegraphics[width=\textwidth]{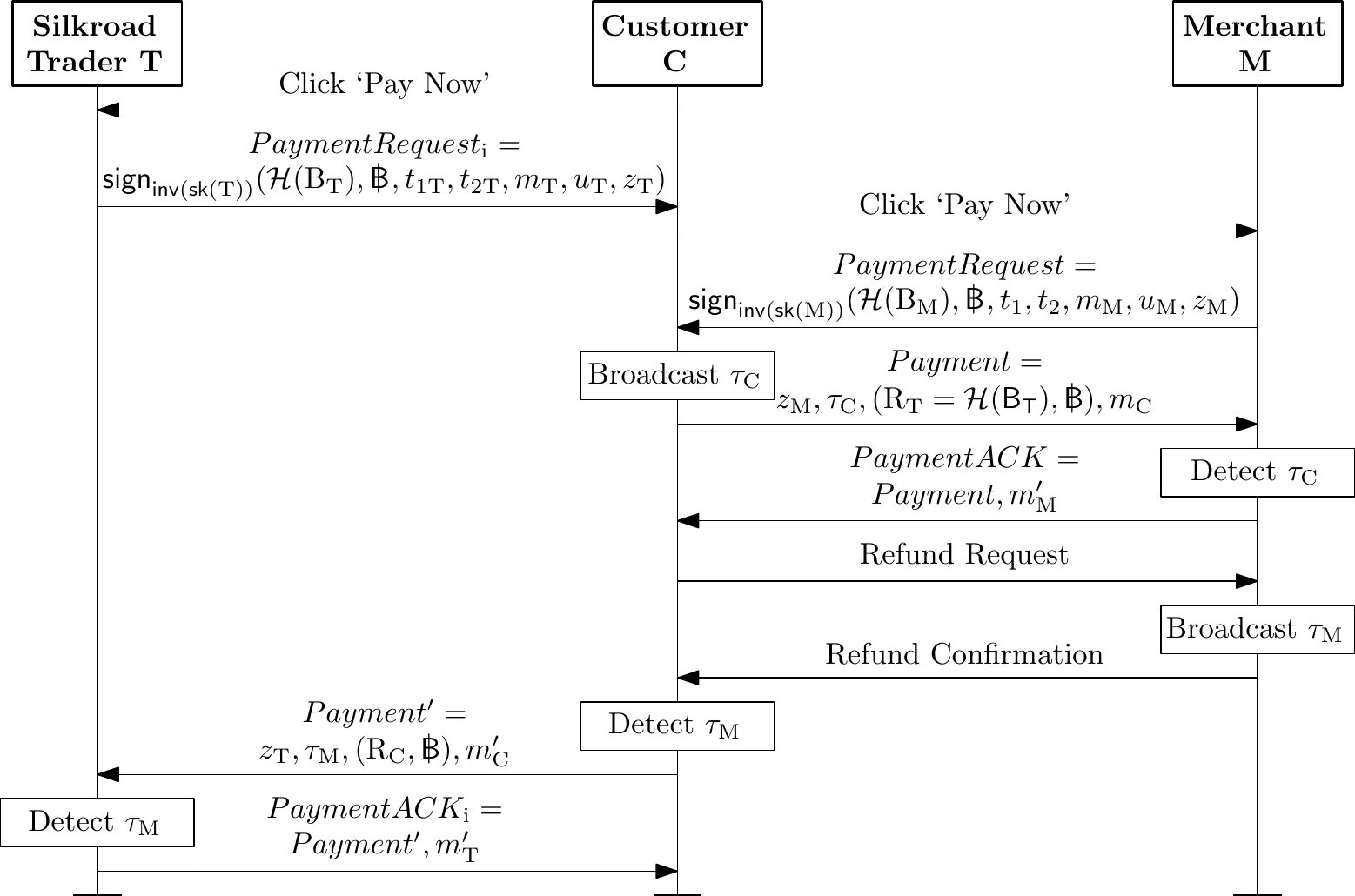}
\caption{\label{fig:attackmessages}The \textit{Silkroad Trader Attack} allows
a customer to route bitcoins to an illicit trader via an honest merchant and then plausibly deny their involvement.
This is achieved by requesting the refund to be issued to the illicit trader. Note that the customer uses the trader's address $\hashsymbol(\bitcoinAddress_{\trader})$ as the refund address within the \emph{Payment} message sent to the merchant.
All messages are sent over an HTTPS communication channel. The notation is specified in Section~\ref{subsec:ModellingBIP70}
and the detailed explanation of the attack can be found in Section~\ref{subsec:Silkroad-Trader-Attack}.}
\end{figure*}

The attack sequence diagram is shown in Figure~\ref{fig:attackmessages}.
Assume a Customer wishes to buy some illicit goods from a `Silkroad Trader'.
The Customer receives a \emph{Payment Request} from the Silkroad Trader ($\trader$) that includes the Silkroad Trader's Bitcoin address $\hashsymbol(\bitcoinAddress_{\trader})$.%
The Customer then finds an honest Merchant supporting the Payment Protocol and selling an item of similar (or possibly just greater) price.
The customer then expresses their wish to buy the item and receives a \emph{Payment Request} from the Merchant.
The Customer then puts together a \emph{Payment} message that includes the payment transaction $\transaction_{\alice}$, but crucially states the Silkroad Trader's address as the refund address, i.e. $\refundAddress_{\trader} = \hashsymbol(\bitcoinAddress_{\trader})$, and sends the \emph{Payment} message to the Merchant.
After finalising the payment and receiving the \emph{Payment Acknowledgement} and before the Merchant ships the item, the Customer cancels the order and requests a refund.
This will prompt the Merchant to prepare and broadcast a refund transaction $\transaction_{\bob}$ that sends the funds to the Silkroad Trader's Bitcoin address.
The Customer will then be able to detect this transaction and include it in a \emph{Payment} message she composes and sends to the Silkroad Trader.
The Silkroad Trader will then detect the transaction, send the \emph{Payment Acknowledgement} to the Customer and ship the illicit goods.

Note that since the \emph{Payment} message sent from the Customer to the Merchant is not digitally signed, the Merchant will not be able to subsequently prove that it was indeed the Customer that requested the funds to be refunded to the Silkroad Trader's address.
Hence, the Customer will be able to pay for the illicit goods through the honest Merchant without leaving any trace of a direct payment to the Silkroad Trader.

\subsection{Model Checking}\label{formal verificaiton_of attack}
We encoded the model described in Section \ref{sec:Payment-Protocol} in the \emph{AnBx}
language \cite{jisa_2016}, an extension of the \emph{AnB} language supported by OFMC, allowing for macro definitions, functions type signature definition and stricter type-checking.
Using the \emph{AnBx compiler} \cite{anbx2015}, we translated the model in the \emph{AnB} format which is suitable for
verification with OFMC. The source code is available in the Appendix.

We run our tests on a Windows 10 PC, with Intel Core i7 4700HQ 2.40\,GHz
CPU and 16\,GB RAM and we verified the model for a single session
in OFMC in the classic and typed mode. As a result, the model demonstrated
that both authentication and secrecy goals were violated. The attack was found in 2.34 seconds.

The authentication goal $\bob$ \texttt{weakly authenticates} $\alice_{i}$ \texttt{on} $\refundAddress_{\alice_{i}},\bitcoinB_{\alice_{i}}$
states for all customers ($i = 1..n$), the merchant can have a guarantee of endorsment of the refund addresses and amounts.

In particular, the goal is violated because in not possible to verify the non-injective agreement~\cite{Lowe97}
between the construction of $\refundAuth{}_{\alice_{i}}=(\refundAddress_{\alice_{i}},\bitcoinB_{\alice_{i}})$  done by $\alice_{i}$ and the corresponding values received by $\bob$.
This is possible because the customers are not required to endorse the value ($\refundAddress_{\alice_{i}},\bitcoinB_{\alice_{i}}$) using digital signatures. Therefore a compromised or dishonest client can easily manipulate the refund address and perform attacks like the one described in \ref{subsec:Silkroad-Trader-Attack}.

The secrecy goal $\left(\refundAddress_{\alice_{1}},\ldots,\refundAddress_{\alice_{n}}\right)$
\texttt{secret between} $\bob,\alice_{1},\ldots,\alice_{n}$ is also violated.
The definition of secrecy used in our model implies that all members
of the secrecy set know the secret values and agree on these. But
in this case, due to lack of authentication, the customer who
is communicating with the merchant can convince other customers that
the refund address she is using is different from the one sent
to the merchant. For example, \emph{$\refundAddress_{\alice_{2}}$},
the refund address of the second customer, can be easily replaced with a different address by $\alice_{1}$ before being communicated to $\bob$.%

It should be noted that, in general, with the automated verification it is not possible to validate a specific attack trace known a priori, and the analysis usually aims at assessing the absence or presence of at least an attack trace that leads at a violation of a security goal. In particular, in order to verify the protocol, the model-checker OFMC builds a state-transition system, and given the initial configuration, analyses the possible transitions in order to see if any attack state is reachable in presence of an active attacker. Therefore, the presence of a specific attack trace is not automatically confirmed, rather, such automated versification helps decide whether any attack trace is present or absent, where an attack trace is defined as a sequence of steps leading to a violation of a given security goal. However, the absence of any attack trace is the ultimate goal of verification, to state the security of the analysed protocol, and we will see how this can be achieved in Section \ref{sec:solution}.

\subsection{Real-World Experiments}

 Our experiments, originally reported in ~\cite{mccorry2016}, aimed to verify the practice of processing refunds by merchants, and assess the feasibility of the attacks. 
 We purchased items from real-world merchants using a modified Bitcoin wallet before requesting for the order to be cancelled and a refund processed. The merchants used during these experiments are based in the UK and are supported by BitPay or Coinbase. 
 The bitcoins used for the experiments are owned by the authors and no money is sent to any illicit trader. 
 All experiments were ethically approved by the relevant research ethics committee.

 \subsubsection{Proof of Concept Wallet}

 We developed a wallet supporting the Payment Protocol and
 automating the \textit{Silkroad Trader} attack. Our wallet works as follows:
 \begin{enumerate}
 \item The customer inserts the illicit trader's \textit{Payment Request}
 URI into the wallet which stores both the request and Bitcoin address
 for later use.
 \item The customer finds an item equal (or greater) in value as the `illicit
 goods' and inserts the merchant's \textit{Payment Request} URI into
 their wallet.
 \item The wallet provides a list of refund addresses that can be chosen
 for the \textit{Payment} message that is sent to the merchant and
 the customer can choose the illicit trader's Bitcoin address.
 \item Assuming a refund has been authorised by the merchant, the wallet
 can detect the merchant's refund transaction on the network and include
 it in a \textit{Payment} message that is sent to the illicit trader.
 \item The wallet is notified by a \textit{Payment Acknowledgement} message
 from the illicit trader that the payment has been received.
 \end{enumerate}
 
 \subsubsection{Simulation of the Attack}

 We discuss the results of carrying out a simulation of the \emph
 \emph{Silkroad Trader} attack against real-world merchants using arbitrary identities, e.g. names and e-mail addresses, created for the experiments only. 

 \emph{Cex} refunded the bitcoins within 3 hours of cancelling the
 order and used the refund address from the Payment Protocol.

 \emph{Pimoroni Ltd.} refunded the bitcoins within a single business
 day and used the refund address from the Payment Protocol.

 \emph{Scan} refunded the bitcoins after 26 days and used the refund
 address from the Payment Protocol. The delay was due to Scan initially
 requesting us to provide a refund address over e-mail, but we insisted
 using the one specified in the original payment message.

 \emph{Dell} were unable to process the refund due to `technical
 difficulties' and requested our bank details. We informed them that
 we did not own a bank account and Dell suggested sending the refund
 as a cheque. While not the experiment's aim, this potentially opens
 Dell as an exchange for laundering tainted bitcoins.

 \subsubsection{Payment Processors' Responses}
 We privately disclosed our attacks to the Payment Processors and received
 the following responses:

 \emph{BitPay} acknowledged ``the researchers have done their homework''
 and that ``refunds are definitely a significant money laundering
 attack vector''. They are now actively monitoring for
 refund activity on behalf of their merchants.

 \emph{Coinbase} acknowledged the \emph{Silkroad Trader} attack as a good example of an authentication vulnerability in the Payment Protocol.

 \emph{Bitt} acknowledged both attacks and  believe the endorsement evidence
 may support Payment Processors to become more `airtight' for future regulation.

 In response to our disclosures, the payment processors have put in place temporary mitigation measures such as monitoring refunds. 
 These measures only partially address the \emph{Silkroad Trader} attack. 
 To fully address the vulnerability, the BIP70 standard would need to be revised, as we have discuss in Section~\ref{sec:solution}.

\section{Verification of the Proposed Solutions\label{sec:solution}}

In our previous work (McCorry et al.~\cite{mccorry2016}), we proposed a solution that requires the refund addresses in the \emph{Payment} message to be endorsed by the customer.
This provides the merchant with a proof of endorsement that can be used to demonstrate to a third party that the customer who authorised the payment also endorsed the refund addresses.
We explain the solution first and then verify, by model-checking, that the solution meets the security goals set out in Section~\ref{subsec:Security-Goals}.
We also present an alternative solution meeting the security goals and discuss the comparative merits of each solution.
\vspace{-2.5mm}

\subsection{The Original Solution}
\label{sub:solution}
\begin{figure*}[t]
\centering
\includegraphics[scale=0.9]{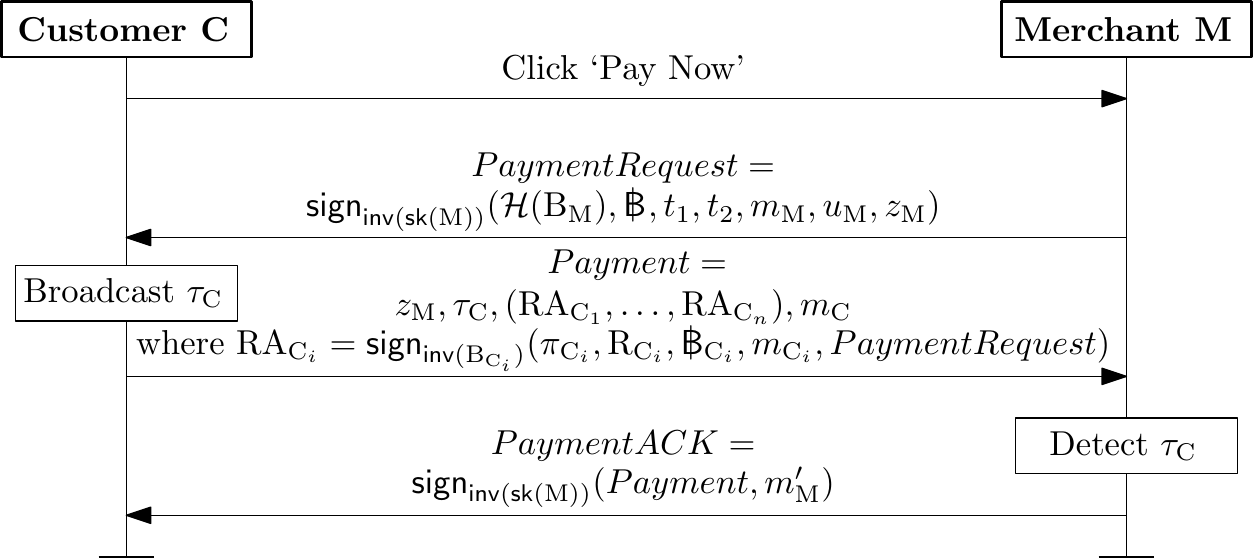}
\caption{\label{fig:alicesig}The Original Solution: expanded message contents for the amended Payment Protocol for Customer $\alice$ and Merchant $\bob$. The Customer endorses the refund addresses by providing digital signatures $\refundAuth_{\alice_{i}}$. The endorsement can be used by the Merchant to prove that they acted as per the instructions of the Customer in case of a refund. The protocol is explained in detail in Section~\ref{sub:solution}.}
\end{figure*}

Figure~\ref{fig:alicesig} shows the amended Payment Protocol as proposed in our earlier work~\cite{mccorry2016}.
The \emph{Payment Request} is similar to before, except that this solution requires the memo $m_{\bob}$ to include specific information to assure the customer that the message is in response to their `click'.
This is to mitigate the Marketplace Trader attack and does not change the formal model we propose in this work.
The \emph{Payment} message includes all the elements specified in the Payment Protocol plus a digital signature to endorse each refund address.
More specifically, for each refund address $\refundAddress_{\alice_{i}}$ and refund amount $\bitcoinB_{\alice_{i}}$, the customer provides a digital signature on $ (R_{\alice_{i}},\bitcoinB_{\alice_{i}},m_{\alice_{i}},\payreq)$ as a proof of endorsement of the refund address and amount that binds these values to the corresponding payment transaction input and the specific $\payreq$.
The \emph{Payment Acknowledgement} contains a copy of the \emph{Payment} message and a memo as before, plus a digital signature by the merchant using the private key corresponding to their X.509 certificate public key.

More specifically, each refund address endorsement signature is in the form:
\[
  \refundAuth_{\alice_{i}}=\sign{\inv{\bitcoinAddress_{\alice_{i}}}}{\pi_{\alice_{i}},\refundAddress_{\alice_{i}},\bitcoinB_{\alice_{i}},m_{\alice_{i}},\payreq},
\]
where $\sign{}{}$ is the signature algorithm and $\inv{\bitcoinAddress_{\alice_{i}}}$ is the private key that authorised the transaction input $\pi_{\alice_{i}}$.
These parameters were chosen to clarify the correspondence between the transaction inputs and the endorsed refund addresses and to ensure the endorsement is only valid for a specific \textit{Payment Request}.
Moreover, the proposed solution suggests that the Merchant should digitally sign the
\emph{Payment Acknowledgement} as follows:
\[
  PaymentACK=\sign{\inv{\sk{\bob}}}{Payment,m'_{\bob}} ,
\]
such that the customers can have evidence of the completion of the protocol.

\paragraph*{Verification}

We updated the model described in Section~\ref{sec:Payment-Protocol} with the new specification, ran the tests and verified the model for a single session in OFMC in the classic and typed mode. With such amendments, we verified in 10.08 seconds that there are no attacks on the security goals for the one-session verification. We also tested the model for two parallel sessions, as long as the available RAM (32 GB) allowed for it. We did not find further attacks.

The fix works because now the refund address and the amount are digitally signed by the private key that authorised the transaction input $\pi_{\alice_{i}}$.
It is now possible to prove the non-injective agreement, i.e. the weak authentication goal.
Since the refund address is endorsed and the merchant receives evidence of endorsement of the refund addresses,
it is now impossible for a compromised or dishonest customer to manipulate these values.

We furthermore tested the stronger authentication goal $\bob$ \texttt{authenticates} $\alice_{i}$ \texttt{on} $\refundAddress_{\alice_{i}},\bitcoinB_{\alice_{i}}$
(for all $i=1..n$ ) and we found it also holds.
This in practice gives the recipients of the signed message evidence of freshness, provided the message contains sufficient information to distinguish between different transactions, which is usually the case.

The secrecy goal is also satisfied because now all the customers and the merchant agree on the refund addresses.
It is not possible for an attacker to manipulate the refund addresses since the digital signatures would not verify in case of manipulation.

In conclusion, we verified that the proposed solution fixes the protocol in that it successfully prevents attacks on the authentication and secrecy goals.

\subsection{An Alternative Solution}
\label{sec:alternative}
\begin{figure*}[t]
\centering
\includegraphics[scale=0.9]{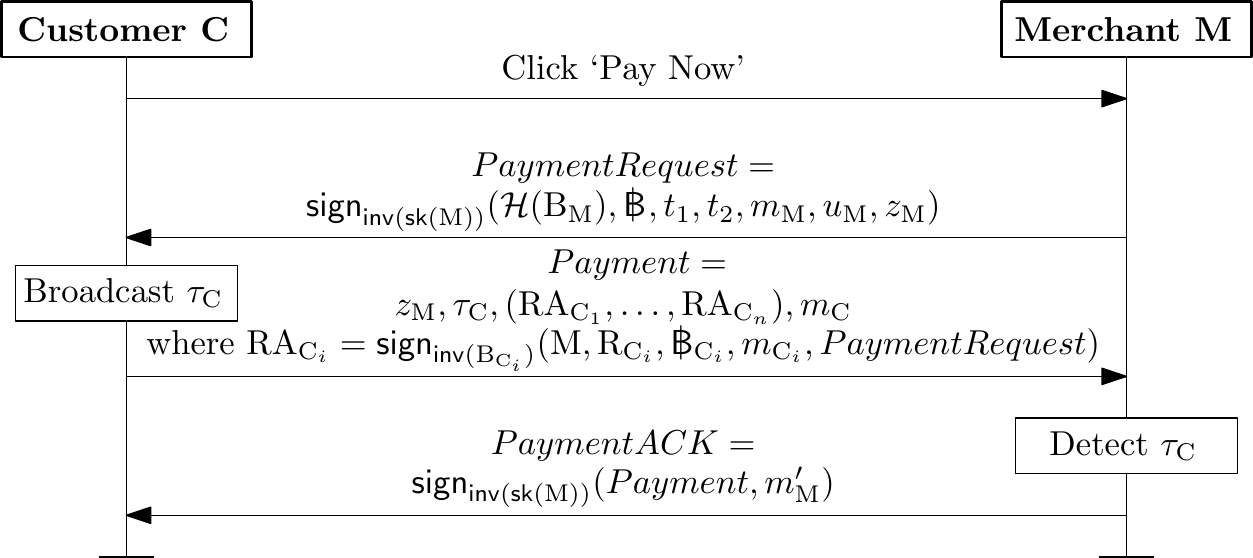}
\caption{\label{fig:alternativesol}Alternative solution: expanded message contents for the amended Payment Protocol for Customer $\alice$ and Merchant $\bob$. The Customer endorsement $\refundAuth_{\alice_{i}}$ does not require the transaction $\pi_{\alice_{i}}$ and instead includes the identity of the Merchant $\bob$. The protocol is explained in detail in Section~\ref{sec:alternative}.}
\end{figure*}

We also propose an alternative solution, depicted in Figure~\ref{fig:alternativesol},
where now the following definition is used:
\[
  \refundAuth{}_{\alice_{i}}=\sign{\inv{\sk{\bitcoinAddress{}_{\alice_{i}}}}}{\bob,\refundAddress_{\alice_{i}},\bitcoinB_{\alice i},m_{\alice_{i}},\payreq} .
\]

In this solution, instead of including $\pi_{\alice_{i}}$,
which introduces a dependency on previous transactions, we simply add the identity of the merchant $\bob$,
which would be immediately available, allowing $\tau_{\alice_{i}}$ to be computed later.
Similarly to the previous solution, all goals, including strong authentication, are successfully verified, in this case significantly faster (5.30 sec).

It is important to note that the authentication goal is met as well because the identity $\bob$ is included in the signed messages and this provides a clear evidence of the customers' intention to run the Payment Protocol with the explicitly identified merchant $\bob$.
Specifically, from the theoretical point of view, this satisfies the definition of the notion of \emph{`agreement'}~\cite{Lowe97} usually considered in formal methods.
This notion is stronger than the kind of guarantee that can be achieved with a digital signature if the recipient identity is omitted in the payload (i.e.\ what is called \emph{`proof of endorsement'}). In practice, binding the refund details to the merchant identity provides an extra guarantee (compared to~\cite{mccorry2016}) that the endorsed refund details can not be reused with merchants other than $\bob$.

\subsection{Discussion}
The main idea behind the proposed solutions is to augment the protocol with a \emph{proof of endorsement} of the refund addresses by the keys that authorise the original transaction.
The merchant may store this proof of endorsement and demonstrate their appropriate conduct to a third party in case of a \textit{Silkroad Trader} attack.
The customer colluding with an illicit trader on the other hand will lose their ability to plausibly deny their involvement in such an attack.
Note that the proof of endorsement only links pseudonymous Bitcoin addresses.
This is the same as what the merchant would learn from the original protocol messages.

It should be pointed out that the proposed extension BIP75~\cite{bip75:payment}, aims at providing the merchant with a publicly verifiable audit log of all transactions.
The solutions considered in this paper also aim to provide the merchant with this audit log but without the need to know the customer's real-world identity. In our approach, customers do not need to maintain certificates to spend their bitcoins. Instead, we propose using the same keys that authorised the Bitcoin transaction to provide the merchant with publicly verifiable evidence.
This evidence states that the same pseudonymous customer that authorised the payment has agreed to the terms of the purchase, and authenticates any new instructions provided by the pseudonymous customer. Every new payment authenticates a new pseudonymous customer and the merchant does not necessarily need to know their identity, but just cares that it is dealing with the correct pseudonymous customer for each payment.

\subsection{Solution Performance}
\label{sec:performance}
The computational overhead of the solution is quite low in general.
On the customer side, the wallet already needs to carry out several cryptographic operations for participating in the original Payment Protocol, including verifying the merchant's certificate chain and signature, signing the payment transaction, and generating refund keys.
Hence, an extra signature (per input) introduces minimal overhead on the customer side. On the merchant side, the overhead is more substantial, however it is limited to the verification of an extra signature and the production of an extra signature for the \emph{Payment Acknowledgement}.

 We have implemented the modifications to the Payment Protocol and measured the performance of all the steps of both the original and the amended Payment Protocols.
 All tests were carried out on a MacBook Pro running OS~X~10.9.1
 with 2.3~GHz Intel Core~i7 and 16~GB DDR3 RAM.
 The details of the measured timing performance can be found in  Table~\ref{tab:timeresults}.
 The reported measurements are for the Bitcoin Core Client while utilising 1 core.
 Furthermore, both signing operations in steps 3 and 8, and the verification operation in step 9, are performed using the Secp256k1 implementation which has replaced OpenSSL in Bitcoin Core~\cite{secp256}.
 Each step was executed 100 times and the reported times represent the average.

 \begin{table*}[t]
 \begin{centering}
 \begin{tabular}{c}
 \textbf{Customer in the current protocol} \tabularnewline
 \begin{tabular}{|p{0.05\textwidth}@{}p{0.7\textwidth}@{\ \ \ }>{\raggedleft\arraybackslash}b{0.15\textwidth}|}
 \hline
 1  & Verify authenticity of merchant's certificate chain & 0.83 ms \tabularnewline
 2  & Verify merchant's signature on Payment Request & 0.08 ms \tabularnewline
 3  & Sign a single transaction input  & 0.08 ms \tabularnewline
 4a  & Fetch list of pre-generated refund addresses $R_{\alice_{1}},...,R_{\alice_{n}}$  & 0.72 ms \tabularnewline
 4b  & Generate new refund address $R_{\alice}$ from wallet key pool  & 110.55 ms \tabularnewline
 5  & Update wallet address book with refund address $R_{\alice}$  & 72.68 ms\tabularnewline \hline
  & \textit{Total without 4b:}  & \textit{74.39 ms}\tabularnewline
  & \textit{Total with 4b:}  & \textit{184.94 ms}\tabularnewline
 \hline \end{tabular} \\
 \rule{0ex}{3ex}
 \textbf{Additional changes proposed for the customer} \tabularnewline
 \begin{tabular}{|p{0.05\textwidth}@{}p{0.7\textwidth}@{\ \ \ }>{\raggedleft\arraybackslash}b{0.15\textwidth}|}
 \hline
 6  & Compute endorsement signature $\refundAuth_{\alice}$ %
 & 0.11 ms\tabularnewline \hline
  & \textit{New Total without 4b:}  & \textit{74.49 ms} \tabularnewline
  & \textit{New Total with 4b:}  & \textit{185.04 ms} \tabularnewline
 \hline \end{tabular} \\
 \rule{0ex}{3ex}
 \textbf{Merchant in the current protocol} \tabularnewline
 \begin{tabular}{|p{0.05\textwidth}@{}p{0.7\textwidth}@{\ \ \ }>{\raggedleft\arraybackslash}b{0.15\textwidth}|}
 \hline
 7  & Verify customer's payment transaction  & 0.29 ms \tabularnewline \hline
  & \textit{Total: }  & \textit{0.29 ms} \tabularnewline
 \hline \end{tabular} \\
 \rule{0ex}{3ex}
 \textbf{Additional changes proposed for the merchant} \tabularnewline
 \begin{tabular}{|p{0.05\textwidth}@{}p{0.7\textwidth}@{\ \ \ }>{\raggedleft\arraybackslash}b{0.15\textwidth}|}
 \hline
  8 & Fetch referenced transaction output & 0.01 ms\tabularnewline
 9  & Verify endorsement signature $\refundAuth_{\alice}$  & 0.13 ms\tabularnewline \hline
  & \textit{New Total: }  & \textit{0.43 ms} \tabularnewline
 \hline
 \end{tabular}
 \end{tabular}
 \par\end{centering}
 \caption{\label{tab:timeresults}Time performance for proposed changes to the
 Payment Protocol}
 \vspace{-2.5mm}
 \end{table*}

 Steps 1\textendash 5 represent the customer's perspective in the current
 Payment Protocol's implementation. The wallet verifies the merchant's
 certificate authenticity using the chain of certificates that lead
 to a trusted root authority and verifies the merchant's signature
 on the \textit{Payment Request} message before authorising at least
 one transaction input to authorise the payment. Then, the wallet fetches
 a list of pre-generated refund addresses and Step 4b only occurs if
 this list is empty as a new refund address must be generated. This
 refund address is associated with the payment for future reference.
 These steps require 74.39~ms if the list of pre-generated refund addresses
 is not empty, otherwise 184.94~ms is required. Our proposed change
 in Step 6 takes 0.11~ms and requires the customer's wallet to sign
 an endorsement message for the refund address, obtaining the signature
 $\sigma_{\alice}$. In total, the time required for the customer is
 185.04~ms with Step 4b, and 74.49~ms without Step 4b.

 Step 7 represents the merchant's perspective in the current Payment
 Protocol's implementation and requires 0.29~ms to check if the payment
 transaction with a single input is valid. We propose in Steps 8\textendash 9
 that the merchant fetches the transaction output referenced in the
 payment transaction's input to let the merchant check the number of
 bitcoins associated with each refund address. Then, the transaction
 input's public key $C$ is used to verify the endorsement signature.
 These proposed changes require 0.14~ms, and in total the time required
 for the merchant is 0.43~ms.

 \vspace{-2.5mm}

\section{Conclusion\label{sec:Conclusion}}
In this work, we considered Bitcoin's Payment Protocol and its vulnerability to an empirically demonstrated attack that leverages the lack of refund address authentication in the protocol to allow malicious customers to route funds through honest merchants to illicit traders and be able to later deny doing so.
We formally modelled the protocol, proved it is vulnerable and validated the solution proposed previously to fix the protocol. We also presented and verified a new alternative solution which is simpler and can have, in principle, a reduced computational impact than the previous one.
In both cases, the solutions provide the merchant with evidence that the refund address
received during the Payment Protocol has been digitally signed from the same pseudonymous
customer who authorised the transaction.

To the best of our knowledge, our model of the Bitcoin Payment Protocol, which complements previous work focusing on underlying core aspects of blockchain and Bitcoin, is the first attempt to formally model and analyse the security of a protocol relaying on Bitcoin. It is worth noting that to complete the analysis we employed specific notions available in OFMC, such as pseudonyms channels and channel abstractions conveying security goals, that allowed us to specify a model that is tractable and can be analysed more efficiently.

\subsection*{Acknowledgements}

The second and forth authors were supported by the European Research Council (ERC) Starting Grant (No. 306994).

\bibliographystyle{abbrv}
\bibliography{literature-refundattacks,literature,literature-formal}

\begin{thebibliography}{10}

\bibitem{bip70:payment}
G.~Andresen and M.~Hearn.
\newblock {BIP 70: Payment Protocol}.
\newblock {\em Bitcoin Improvement Process}, July 2013.
\newblock \url{https://github.com/bitcoin/bips/blob/master/bip-0070.mediawiki}.

\bibitem{Arapinis2019}
M.~Arapinis, A.~Gkaniatsou, D.~Karakostas, and A.~Kiayias.
\newblock A formal treatment of hardware wallets.
\newblock In Goldberg and Moore, editors, {\em Financial Cryptography and Data
  Security (FC~2019)}, volume 11598 of {\em LNCS}, pages 426--445. Springer,
  2019.

\bibitem{Atzei2018}
N.~Atzei, M.~Bartoletti, S.~Lande, and R.~Zunino.
\newblock A formal model of bitcoin transactions.
\newblock In Meiklejohn and Sako, editors, {\em Financial Cryptography and Data
  Security (FC~2018)}, volume 10957 of {\em LNCS}, pages 541--560. Springer,
  2018.

\bibitem{AvizhehSS18}
S.~Avizheh, R.~Safavi{-}Naini, and S.~F. Shahandashti.
\newblock {A New Look at the Refund Mechanism in the Bitcoin Payment Protocol}.
\newblock In {\em Financial Cryptography}, volume 10957 of {\em Lecture Notes
  in Computer Science}, pages 369--387. Springer, 2018.

\bibitem{basin2005ofmc}
D.~Basin, S.~M\"odersheim, and L.~Vigan\`o.
\newblock {OFMC}: A symbolic model checker for security protocols.
\newblock {\em International Journal of Information Security}, 4(3):181--208,
  2005.

\bibitem{Bhargavan2017}
K.~Bhargavan, B.~Blanchet, and N.~Kobeissi.
\newblock Verified models and reference implementations for the {TLS} 1.3
  standard candidate.
\newblock In {\em 2017 {IEEE} Symposium on Security and Privacy, {SP} 2017, San
  Jose, CA, USA, May 22-26, 2017}, pages 483--502. {IEEE} Computer Society,
  2017.

\bibitem{bhargavan2016formal}
K.~Bhargavan, A.~Delignat-Lavaud, C.~Fournet, A.~Gollamudi, G.~Gonthier,
  N.~Kobeissi, A.~Rastogi, T.~Sibut-Pinote, N.~Swamy, and S.~Zanella-Beguelin.
\newblock Formal verification of smart contracts.
\newblock In {\em Proceedings of the 2016 ACM Workshop on Programming Languages
  and Analysis for Security PLAS16}, pages 91--96, 2016.

\bibitem{bitcoin20core}
{Bitcoin Core Team}.
\newblock {Bitcoin Core} release notes 0.20.0, 2020.
\newblock
  \url{https://github.com/bitcoin/bitcoin/blob/master/doc/release-notes/release-notes-0.20.0.md/}.

\bibitem{lightning-usage-stats-bitmex}
{BitMEX Research}.
\newblock Lightning network (part 6) - over 60,000 non-cooperative channel
  closures, 2020.
\newblock
  \url{https://blog.bitmex.com/lightning-network-part-6-over-60000-non-cooperative-channel-closures}.

\bibitem{virus16bitpay}
BitPay.
\newblock Avoiding and detecting a new virus affecting some bitcoin users,
  2016.
\newblock
  \url{https://bitpay.com/blog/avoiding-and-detecting-a-new-virus-affecting-some-bitcoin-users/}.

\bibitem{bitpay18payment}
BitPay.
\newblock How payment protocol is eliminating costly bitcoin payment errors:
  Stats and results, 2018.
\newblock \url{https://bitpay.com/blog/payment-protocol-results}.

\bibitem{jisa_2016}
M.~Bugliesi, S.~Calzavara, S.~M{\"o}dersheim, and P.~Modesti.
\newblock Security protocol specification and verification with {AnBx}.
\newblock {\em Journal of Information Security and Applications}, 30:46--63,
  2016.

\bibitem{chainalysis20}
{Chainalysis Team}.
\newblock Darknet market activity higher than ever in 2019 despite closures.
  how does law enforcement respond?, 2020.
\newblock
  \url{https://blog.chainalysis.com/reports/darknet-markets-cryptocurrency-2019}.

\bibitem{Chaudhary2015}
K.~Chaudhary, A.~Fehnker, J.~van~de Pol, and M.~Stoelinga.
\newblock Modeling and verification of the bitcoin protocol.
\newblock In van Glabbeek, Groote, and H{\"{o}}fner, editors, {\em Proceedings
  Workshop on Models for Formal Analysis of Real Systems, {MARS}~2015}, volume
  196 of {\em {EPTCS}}, pages 46--60, 2015.

\bibitem{Cremers2017}
C.~Cremers, M.~Horvat, J.~Hoyland, S.~Scott, and T.~van~der Merwe.
\newblock A comprehensive symbolic analysis of {TLS} 1.3.
\newblock In B.~M. Thuraisingham, D.~Evans, T.~Malkin, and D.~Xu, editors, {\em
  Proceedings of the 2017 {ACM} {SIGSAC} Conference on Computer and
  Communications Security ({CCS} 2017)}, pages 1773--1788. {ACM}, 2017.

\bibitem{dolev83ieee}
D.~Dolev and A.~Yao.
\newblock On the security of public-key protocols.
\newblock {\em IEEE Transactions on information Theory}, 2(29), 1983.

\bibitem{duan2018formal}
Z.~Duan, H.~Mao, Z.~Chen, X.~Bai, K.~Hu, and J.-P. Talpin.
\newblock Formal modeling and verification of blockchain system.
\newblock In Meiklejohn and Sako, editors, {\em Proceedings of the 10th
  International Conference on Computer Modeling and Simulation}, volume 10957
  of {\em LNCS}, pages 231--235. Springer, 2018.

\bibitem{dwivedi2019formal}
V.~Dwivedi, V.~Deval, A.~Dixit, and A.~Norta.
\newblock Formal-verification of smart-contract languages: A survey.
\newblock In {\em International Conference on Advances in Computing and Data
  Sciences}, pages 738--747. Springer, 2019.

\bibitem{amcu}
C.~Egger, P.~Moreno-Sanchez, and M.~Maffei.
\newblock Atomic multi-channel updates with constant collateral in
  bitcoin-compatible payment-channel networks.
\newblock In {\em Computer and Communications Security (CCS)}, pages 801--815.
  ACM, 2019.

\bibitem{AnBEMV}
L.~Freitas, P.~Modesti, and M.~Emms.
\newblock {A Methodology for Protocol Verification applied to EMV1}.
\newblock In {\em Formal Methods: Foundations and Applications - 21th Brazilian
  Symposium, {SBMF} 2018, Proceedings}, volume 11254 of {\em LNCS}. Springer,
  2018.

\bibitem{garay2015bitcoin}
J.~A. Garay, A.~Kiayias, and N.~Leonardos.
\newblock The bitcoin backbone protocol: Analysis and applications.
\newblock In {\em EUROCRYPT (2)}, pages 281--310, 2015.

\bibitem{Harz2018}
D.~Harz and W.~J. Knottenbelt.
\newblock Towards safer smart contracts: {A} survey of languages and
  verification methods.
\newblock {\em CoRR}, abs/1809.09805, 2018.

\bibitem{overkill}
M.~Hearn.
\newblock {Re: [Bitcoin-development] BIP 70 refund field}.
\newblock {\em Bitcoin-Development}, Mar. 2014.
\newblock \url{{http://sourceforge.net/p/bitcoin/mailman/message/ 32157661/}}.

\bibitem{Herzog2005}
J.~Herzog.
\newblock A computational interpretation of dolev-yao adversaries.
\newblock {\em Theor. Comput. Sci.}, 340(1):57--81, 2005.

\bibitem{Hildenbrandt2018}
E.~Hildenbrandt, M.~Saxena, N.~Rodrigues, X.~Zhu, P.~Daian, D.~Guth, B.~M.
  Moore, D.~Park, Y.~Zhang, A.~Stefanescu, and G.~Rosu.
\newblock {KEVM:} {A} complete formal semantics of the ethereum virtual
  machine.
\newblock In {\em 31st {IEEE} Computer Security Foundations Symposium, {CSF}
  2018}, pages 204--217. {IEEE} Computer Society, 2018.

\bibitem{pp-wallets-bitpay}
{Keaton (BitPay Support)}.
\newblock Which wallets work best for a bitpay payment? {Which} wallets are
  compatible?, June 2020.
\newblock \url{https://support.bitpay.com/hc/en-us/articles/115005701523}.

\bibitem{Klomp2018}
R.~Klomp and A.~Bracciali.
\newblock On symbolic verification of bitcoin's script language.
\newblock In Garc{\'{\i}}a{-}Alfaro, Herrera{-}Joancomart{\'{\i}}, Livraga, and
  Rios, editors, {\em Data Privacy Management, Cryptocurrencies and Blockchain
  Technology - {ESORICS} 2018 International Workshops, {DPM} 2018 and {CBT}
  2018, Proceedings}, volume 11025 of {\em LNCS}, pages 38--56. Springer, 2018.

\bibitem{Lowe97}
G.~Lowe.
\newblock A hierarchy of authentication specifications.
\newblock In {\em CSFW'97}, pages 31--43. IEEE Computer Society Press, 1997.

\bibitem{luu2016making}
L.~Luu, D.-H. Chu, H.~Olickel, P.~Saxena, and A.~Hobor.
\newblock Making smart contracts smarter.
\newblock In {\em Proceedings of the 2016 ACM SIGSAC Conference on Computer and
  Communications Security}, pages 254--269. ACM, 2016.

\bibitem{McCorryMSH16PayNet}
P.~McCorry, M.~M{\"{o}}ser, S.~F. Shahandashti, and F.~Hao.
\newblock Towards bitcoin payment networks.
\newblock In {\em {ACISP} {(1)}}, volume 9722 of {\em Lecture Notes in Computer
  Science}, pages 57--76. Springer, 2016.

\bibitem{McCorrySCH15AKE}
P.~McCorry, S.~F. Shahandashti, D.~Clarke, and F.~Hao.
\newblock {Authenticated Key Exchange over Bitcoin}.
\newblock In {\em {Security Standardisation Research}}, volume 9497 of {\em
  Lecture Notes in Computer Science}, pages 3--20. Springer, 2015.

\bibitem{mccorry2016}
P.~McCorry, S.~F. Shahandashti, and F.~Hao.
\newblock Refund attacks on bitcoin`s payment protocol.
\newblock In {\em 20th Financial Cryptography and Data Security conference},
  2016.

\bibitem{miller2019sprites}
A.~Miller, I.~Bentov, S.~Bakshi, R.~Kumaresan, and P.~McCorry.
\newblock Sprites and state channels: Payment networks that go faster than
  lightning.
\newblock In {\em Financial Cryptography and Data Security}, pages 508--526.
  Springer, 2019.

\bibitem{anb}
S.~M{\"o}dersheim.
\newblock Algebraic properties in {A}lice and {B}ob notation.
\newblock In {\em International Conference on Availability, Reliability and
  Security (ARES 2009)}, pages 433--440, 2009.

\bibitem{moedersheim2009secure}
S.~M{\"o}dersheim and L.~Vigan{\`o}.
\newblock Secure pseudonymous channels.
\newblock In {\em Computer Security--ESORICS 2009: 14th European Symposium on
  Research in Computer Security, Saint-Malo, France, September 21-23, 2009,
  Proceedings}, page 337. Springer, 2009.

\bibitem{Moedersheim2014}
S.~M\"{o}dersheim and L.~Vigan\`{o}.
\newblock Sufficient conditions for vertical composition of security protocols.
\newblock In {\em Proceedings of the 9th ACM Symposium on Information, Computer
  and Communications Security}, ASIA CCS '14, pages 435--446, New York, NY,
  USA, 2014. ACM.

\bibitem{anbx2015}
P.~Modesti.
\newblock {AnBx}: Automatic generation and verification of security protocols
  implementations.
\newblock In {\em 8th International Symposium on Foundations \& Practice of
  Security}, volume 9482 of {\em LNCS}. Springer, 2015.

\bibitem{nakamoto2008bitcoin}
S.~Nakamoto.
\newblock {Bitcoin: A Peer-to-Peer Electronic Cash System}, November 2008.
\newblock \url{https://bitcoin.org/bitcoin.pdf}.

\bibitem{bip75:payment}
J.~Newton, M.~David, A.~Voisine, and M.~James.
\newblock {BIP 75: Out of Band Address Exchange using Payment Protocol
  Encryption}.
\newblock {\em Bitcoin Improvement Process}, Nov. 2015.
\newblock \url{https://github.com/bitcoin/bips/blob/master/bip-0075.mediawiki}.

\bibitem{OConnor2017}
R.~O'Connor.
\newblock Simplicity: {A} new language for blockchains.
\newblock In {\em Proceedings of the 2017 Workshop on Programming Languages and
  Analysis for Security, PLAS@CCS 2017}, pages 107--120. {ACM}, 2017.

\bibitem{lightning16}
J.~Poon and T.~Dryja.
\newblock The bitcoin lightning network: Scalable off-chain instant payments,
  2016.
\newblock \url{https://lightning.network/lightningnetwork-paper.pdf}.

\bibitem{bip21:payment}
N.~Schneider and M.~Corallo.
\newblock {BIP 21: URI Scheme}.
\newblock {\em Bitcoin Improvement Process}, 2012.
\newblock \url{https://github.com/bitcoin/bips/blob/master/bip-0021.mediawiki}.

\bibitem{lightning-usage-stats-cryptonews}
K.~Torpey.
\newblock Is bitcoin's lightning network ready to replace altcoin use cases?,
  2020.
\newblock
  \url{https://cryptonews.com/exclusives/is-bitcoin-s-lightning-network-ready-to-replace-altcoin-use-6511.htm}.

\bibitem{Tsankov2018}
P.~Tsankov, A.~M. Dan, D.~Drachsler{-}Cohen, A.~Gervais, F.~B{\"{u}}nzli, and
  M.~T. Vechev.
\newblock Securify: Practical security analysis of smart contracts.
\newblock In D.~Lie, M.~Mannan, M.~Backes, and X.~Wang, editors, {\em
  Proceedings of the 2018 {ACM} {SIGSAC} Conference on Computer and
  Communications Security, {CCS} 2018}, pages 67--82. {ACM}, 2018.

\bibitem{Turuani2016}
M.~Turuani, T.~Voegtlin, and M.~Rusinowitch.
\newblock Automated verification of electrum wallet.
\newblock In Clark, Meiklejohn, Ryan, Wallach, Brenner, and Rohloff, editors,
  {\em Financial Cryptography and Data Security - {FC} 2016 International
  Workshops, BITCOIN, VOTING, and WAHC}, volume 9604 of {\em LNCS}, pages
  27--42. Springer, 2016.

\bibitem{malicious14extension}
W.~Wei.
\newblock Malicious chrome extension hijacks cryptocurrencies and wallets,
  2014.
\newblock
  \url{https://thehackernews.com/2014/04/malicious-chrome-extension-hijacks.html}.

\bibitem{secp256}
P.~Wuille.
\newblock {Switch to libsecp256k1-based ECDSA validation}.
\newblock {\em Bitcoin Github Repository}, Nov. 2015.
\newblock \url{https://github.com/bitcoin/bitcoin/pull/6954}.

\end{thebibliography}

\appendix

\section*{Appendix\label{sec:Appendix}}

\begin{lstlisting}[language=AnBx]
Protocol: BIP70 AnB

Types:

    Agent C1,C2,M;
    Number paynow,BTC,BTC1,BTC2,t1,t2,MM,UM,ZM,MM1,RC1,RC2;
    PublicKey BAM,BC1,BC2;
    Function [Agent -> PublicKey] sk;
    Function [Number,Number -> Number] tr;
    Function [Untyped -> Number] hash

Definitions:

    # Bitcoin transaction outputs (previous transaction)
    OutC1: BTC1,hash(BC1);
    OutC2: BTC2,hash(BC2);

    # we ignore inputs as they are not relevant for the next transaction
    TauC1: tr(OutC1);
    TauC2: tr(OutC2);

    # Bitcoin transaction inputs
    PiC1 : {hash(TauC1)}inv(BC1),BC1;
    PiC2 : {hash(TauC2)}inv(BC2),BC2;

    # Bitcoin transaction inputs
    Pi: PiC1,PiC2;

    # Payment request from merchant
    PaymentRequest:{hash(BAM),BTC,t1,t2,MM,UM,ZM}inv(sk(M));

    # SilkRoad attack
    # merchant does not have a guarantee that
    # refund addresses RCx are authorised by BCx

    # <-- UNCOMMENT THESE LINES TO REPRODUCE THE ATTACK
    # and COMMENT the other definition of RACx
    # Refund address authorization (original specification)
    # RAC1: RC1,BTC1;
    # RAC2: RC2,BTC2;

    # McCorry et al. solution: the payment request 
    # is needed to link the refund request to the transaction
    # RAC1: {PiC1,RC1,BTC1,PaymentRequest}inv(BC1);
    # RAC2: {PiC2,RC2,BTC2,PaymentRequest}inv(BC2);
    
    # alternative solution
    RAC1: {M,RC1,BTC1,PaymentRequest}inv(BC1);
    RAC2: {M,RC2,BTC2,PaymentRequest}inv(BC2);

    TauC: Pi,(BTC,hash(BAM))

    # The fix includes a message digitally signed
    # with the session private key of C1 or C2 - inv(BCx)
    # The signed message includes the identities of M and the
    # Refund addresses RCx, the bitcoin amount BTCx, PiCx
    # Note that C1,M are required to achieve the weak authentication goal.
    # In practice C1, M may be an unique identifier on which
    # the parties agree, a public key for example

Knowledge:

    # initial knowledge of agents
    # M does not know C1,C2 - C1,C2 know each other - C2 does not know M
    C2: C2,C1,sk,hash,tr;
    C1: C1,M,C2,sk,hash,paynow,tr;
    M: M,sk,inv(sk(M)),hash,paynow,t1,t2,tr

Actions:

    # the previous transactions are exchanged between C1,C2
    # made public so the intruder can learn them
    # this is a fictional exchange as this information 
    # is already public on the blockchain
    C1 -> C2: TauC1
    C2 -> C1: TauC2

    # starting payment protocol
    [C1] *->* M: paynow

    # payment request signed by the merchant
    M *->* [C1]: PaymentRequest

    # C1,C2 cooperate to construct a transaction
    # (not part of payment protocol)
    C1 *->* C2: RAC1,M,PaymentRequest,BC1
    C2 *->* C1: RAC2,PiC2

    # Payment
    [C1] *->* M: ZM,TauC,RAC1,RAC2

    # Payment Acknowledgement
    # original specification
    # M *->* [C1]: ZM,TauC,RAC1,RAC2,MM1
    
    # McCorry et al. and alternative solution
    M *->* [C1]: {ZM,TauC,RAC1,RAC2,MM1}inv(sk(M))

    # end of the payment protocol

    # agents names are revealed to M
    [C1] *->* M: C1,C2

    # TauC is made public (broadcasted) and learned by the intruder
    M -> C1: TauC

    # Public Keys are made available
    C1 -> M: BC1,BC2,TauC1,TauC2

    M -> C1: BAM
    C1 -> C2: BAM

Goals:

    # M has a guarantee that RCx,BTCx are provided by and linked to Cx
    M weakly authenticates C1 on RC1,BTC1
    M weakly authenticates C2 on RC2,BTC2

    # RCx are confidential
    RC1,RC2 secret between M,C1,C2

    # other goals
    M weakly authenticates [C1] on RC1,BTC1,RC2,BTC2
    M weakly authenticates [C1] on RC1,BTC1
    M weakly authenticates [C1] on BC1
\end{lstlisting}
\end{document}